\documentclass[runningheads,a4paper]{tlp}

\usepackage{amssymb}
\usepackage{amsmath}
\usepackage{lmodern}
\usepackage{verbatim}
\usepackage{graphicx}
\usepackage{caption}
\usepackage{subcaption}
\usepackage{hyperref}
\usepackage{algorithmicx}
\usepackage{algorithm}
\usepackage[noend]{algpseudocode}
\usepackage{enumerate}
\usepackage{multirow}
\usepackage{pdfpages}
\newtheorem{definition}{Definition}
\newtheorem{example}{Example}

\newcommand{\A}{\ensuremath{{\cal A}}}
\newcommand{\AC}{\ensuremath{{\cal A}_{C}}}
\newcommand{\LabAClit}[1]{\ensuremath{\AC^{\#}(#1)}}

\newcommand{\cache}{\ensuremath{\mathbb{M}}}
\newcommand{\upd}[1]{\ensuremath{\textsf{upd}(\theta,\cache,#1)}}
\renewcommand{\P}{\ensuremath{{\cal P}}}

\newcommand{\kbd}[1]{\mbox{\tt #1}}
\newcommand{\skbd}[1]{\mbox{\tt\small{#1}}}

\newcommand{\gd}[0]{\mid}
\newcommand{\state}[2]
  {\ensuremath{\langle #1 \gd{} #2 \rangle}}
\newcommand{\exstate}[3]
  {\ensuremath{\langle #1 \gd{} #2 \gd{} #3 \rangle}}
\newcommand{\exstateM}[4]
  {\ensuremath{\langle #1 \gd{} #2 \gd{} #3 \gd{} #4 \rangle}}

\newcommand{\emptyGoal}{\ensuremath{\square}}

\newcommand{\ADeps}{\ensuremath{{\cal E}}}

\newcommand{\defn}[1]
  {\ensuremath{\textsf{defn}(#1)}}

\newcommand{\answers}[1]{\ensuremath{\textsf{answers}(#1)}}

\newcommand{\clause}[2]
  {\ensuremath{#1\textsf{~:-~} #2}}

\newcommand{\succTriv}[2]
  {\ensuremath{#1 \Rightarrow_P #2}}

\newcommand{\succTrivCache}[2]
  {\ensuremath{#1 \overset{\cache}{\Rightarrow}_P #2}}
\newcommand{\notSuccTrivCache}[2]
  {\ensuremath{#1 \overset{\cache}{\not\Rightarrow}_P #2}}

\newcommand{\reduction}[2]
  {\ensuremath{#1 \leadsto #2}}
\newcommand{\reductionStar}[2]
  {\ensuremath{#1 \leadsto^* #2}}

\newcommand{\reductionA}[2]
  {\ensuremath{#1 \leadsto_{\A} #2}}

\newcommand{\asrId}[1]{\ensuremath{c_{#1}}}
\newcommand{\negAsrId}[1]{\ensuremath{\bar{c}_{#1}}}

\newcommand{\checkLitLab}[2]{\ensuremath{\textsf{check}(#1, #2)}}

\newcommand{\callsAsr}[2]{\ensuremath{\textsf{calls}(#1, #2)}}

\newcommand{\successAsr}[3]{\ensuremath{\textsf{success}(#1, #2, #3)}}

\newcommand{\labCallsAsr}[3]
  {\ensuremath{\textsf{\asrId{#1}\#calls}(#2, #3)}}
\newcommand{\labSuccessAsr}[4]
  {\ensuremath{\textsf{\asrId{#1}\#success}(#2, #3, #4)}}

\submitted{April 29, 2015}
\revised{July 3, 2015}
\accepted{July 14, 2015}

\title[Practical Run-time Checking via Unobtrusive Property Caching]
      {Practical Run-time Checking\\ via Unobtrusive Property Caching}

\author[Nataliia Stulova, Jos\'{e} F. Morales,
  and Manuel V. Hermenegildo]{
       NATALIIA STULOVA$^{1}$
    ~~ JOS\'{E} F. MORALES$^{1}$
    ~~ MANUEL V. HERMENEGILDO$^{1,2}$
\ \\
\ \\
   $^1$IMDEA Software Institute \\
   \email{\{nataliia.stulova, josef.morales,
     manuel.hermenegildo\}@imdea.org} \vspace*{1mm} \\
   $^2$School of Computer Science, Technical University of Madrid (UPM) \\
   \email{manuel.hermenegildo@upm.es}\\
}

\begin{document}
\maketitle

\begin{abstract}
  The use of annotations, referred to as assertions or contracts, to
  describe program properties for which run-time tests are to be
  generated, has become frequent in dynamic programing languages.
  However, the frameworks proposed to support such run-time testing
  generally incur high time and/or space overheads over standard
  program execution.
  We present an approach for reducing this overhead that is based
  on the use of memoization to cache intermediate results of check
  evaluation, avoiding repeated checking of previously verified
  properties.
  Compared to approaches that reduce checking frequency, our proposal
  has the advantage of being exhaustive (i.e., all tests are checked
  at all points) while still being much more efficient than standard
  run-time checking. Compared to the limited previous work on
  memoization, it performs the task without requiring
  modifications to data structure representation or checking code.
  While the approach is general and system-independent, we present it
  for concreteness in the context of the Ciao run-time checking
  framework, which allows us to provide an operational semantics
  with checks and caching.
  We also report on a prototype implementation and provide some
  experimental results that support that using a relatively small cache
  leads to significant decreases in run-time checking overhead.
  To appear in Theory and Practice of Logic Programming (TPLP), Proceedings of ICLP 2015.
\end{abstract}
\vspace{-2em}


\section{Introduction}
\label{sec:intro}

The use of annotations to describe program properties for which
run-time tests are to be generated has become frequent in dynamic
programming languages, including assertion-based approaches in
(Constraint) Logic Programming ((C)LP)~\cite{DNM88-short,%
assert-lang-ws,aadebug97-informal-short,%
BDM97,prog-glob-an,assrt-theoret-framework-lopstr99,
DBLP:conf/discipl/Lai00,ciaopp-sas03-journal-scp-short,%
testchecks-iclp09},
soft/gradual typing in functional
programming~\cite{cartwright91:soft_typing-short,DBLP:conf/icfp/FindlerF02,%
TypedSchemeF08-short,DBLP:journals/toplas/DimoulasF11},
and contract-based extensions in object-oriented
programming~\cite{lamport99:types_spec_lang,clousot-2010,%
DBLP:journals/fac/LeavensLM07}.
However, run-time testing in these frameworks can generally incur high
penalty in execution time and/or space over the standard program
execution without tests.  A number of techniques have been proposed to
date to reduce this overhead, including simplifying the checks at
compile time via static
analysis~\cite{assert-lang-ws,aadebug97-informal-short,prog-glob-an}
or reducing
the frequency of checking, including for example testing only at a
reduced number of points~\cite{testchecks-iclp09,profiling-padl11}.

Our objective
is to develop an approach to run-time
testing that is efficient while being minimally obtrusive and
remaining exhaustive. We present an approach
based on the use of memoization to cache intermediate results of check
evaluation in order to avoid repeated checking of previously verified
properties over the same data structure.
Memoization has of course a long tradition in (C)LP in uses such as
tabling
resolution~\cite{tamaki.iclp86-short,%
dietrich87:_exten_tables_recur_query_evaluat,Warren92-short},
including also sharing and memoizing tabled
sub-goals~\cite{hash-consing}, for improving termination. Memoization
has also been used in program analysis~\cite{pracabsin,ai-jlp}, where
tabling resolution is performed using abstract values.  However,
in tabling and analysis what is tabled are call-success patterns and in
our case the aim is to cache the results of test execution.

While the approach that we propose is general and system-independent,
we will present it for concreteness in the context of the Ciao
run-time checking framework.
The Ciao model~\cite{prog-glob-an,assrt-theoret-framework-lopstr99,%
  ciaopp-sas03-journal-scp-short}
is well understood, and different aspects of it
have been incorporated in popular (C)LP systems, such as Ciao, SWI, and
XSB~\cite{hermenegildo11:ciao-design-tplp,%
  xsb-journal-2012,DBLP:journals/tplp/MeraW13}.
Using this concrete model allows us to provide an operational
semantics of programs with
checks and caching,
as well as a concrete implementation from which we derive experimental
results.
We also present a
program transformation for implementing
the run-time checks that is more efficient than previous
proposals~\cite{assrt-theoret-framework-lopstr99,testchecks-iclp09,profiling-padl11}.
Our experimental results provide evidence that using a relatively
small cache leads to significant decreases in run-time checking overhead.


\section{Preliminaries}
\label{sec:prel-notat}

\paragraph{Basic notation and standard semantics.}

We recall some concepts and notation from standard (C)LP theory.
An \emph{atom} has the form $p(t_1,...,t_n)$ where
$p$ is a predicate symbol of arity $n$ and $t_1,...,t_n$ are terms.
A \emph{constraint} is
a conjunction of expressions built
from predefined predicates (such as term equations or inequalities
over the reals) whose arguments are constructed using predefined
functions (such as real addition).
A \emph{literal} is either an atom or a constraint.
A \emph{goal} is a finite sequence of literals.
A \emph{rule} is of the form $H \mbox{\tt :-} B$ where $H$, the
\emph{head}, is an atom and $B$, the \emph{body}, is a possibly empty
finite sequence of literals.
A \emph{constraint logic program}, or \emph{program}, is a finite set
of rules.

The \emph{definition} of an atom $A$ in a program, $\defn{A}$, is the
set of variable renamings of the program rules
s.t.\ each renaming
has $A$ as a head and has distinct new local variables.
We assume that all rule heads are \emph{normalized}, i.e., $H$ is of the
form $p(X_1,...,X_n)$ where the $X_1,...,X_n$ are distinct free
variables.
Let $\overline{\exists}_L \theta$ be the constraint $\theta$ restricted
to the variables of the syntactic object $L$.
We denote
\emph{constraint entailment} by $\models$, so that $\theta_1\models
\theta_2$ denotes that $\theta_1$ entails $\theta_2$.
Then, we say that $\theta_2$ is \emph{weaker} than $\theta_1$.


The operational semantics of a program is given in terms of its
``derivations'', which are sequences of ``reductions'' between ``states''.
A \emph{state} $\state{G}{\theta}$ consists of a goal $G$ and a
constraint store (or \emph{store} for short) $\theta$.
We use :: to denote concatenation of sequences and we assume for
simplicity that the underlying constraint solver is complete.
A state $S = \state{L::G}{\theta}$ where $L$ is a literal can be
\emph{reduced} to a state $S'$ as follows:
\vspace*{-2mm}
\begin{enumerate}
\item $\reduction{\state{L::G}{\theta}}{\state{G}{\theta \land L}}$
  if $L$ is a constraint and $\theta \land L$ is satisfiable.
\item $\reduction{\state{L::G}{\theta}}{\state{B::G}{\theta}}$ if
  $L$ is an atom of the form $p(t_1,\ldots,t_n)$,\\ for some rule $(L
  \mbox{\tt :-} B)$ $\in \defn{L}$.
\end{enumerate}
\vspace*{-2mm}
We use $\reduction{S}{S'}$ to indicate that a reduction can be applied
to state $S$ to obtain state $S'$. Also, $\reductionStar{S}{S'}$
indicates that there is a sequence of reduction steps from state $S$
to state $S'$.
A \emph{query} is a pair $(L,\theta)$, where $L$ is a literal and
$\theta$ a store, for which the (C)LP system starts a computation from
state $\state{L}{\theta}$.
A finished derivation from a query $(L,\theta)$ is
\emph{successful} if the last state is of the form
$\state{\emptyGoal}{\theta'}$, where $\emptyGoal$ denotes the empty
goal sequence. In that case, the constraint $\bar{\exists}_{L}\theta'$
is an \emph{answer} to $S$.
We denote by $\answers{Q}$ the set of answers to a query $Q$.

\paragraph{\kbd{pred}-Assertions and their Semantics.}
Assertions are linguistic constructions for expressing properties of
programs.
Herein, we will use the \kbd{pred}-assertions
of~\cite{prog-glob-an,assert-lang-disciplbook-short,assrt-theoret-framework-lopstr99},
for which we follow the formalization of~\cite{asrHO-ppdp2014}.
These assertions allow specifying certain conditions on the constraint
store that must hold at certain points of program derivations.
In particular, they allow stating sets of
\emph{preconditions} and \emph{conditional postconditions} for a
given predicate.
A set of assertions for a predicate is of the form:\\ [-8mm]

\begin{small}
  \[
  \begin{array}{l}
    \kbd{:- pred } Head \kbd{ : } Pre_1 \kbd{ => } Post_1 \kbd{.}
  \\
    \ldots
  \\
    \kbd{:- pred } Head \kbd{ : } Pre_n \kbd{ => } Post_n \kbd{.}
  \end{array}
  \]
\end{small}
\hspace*{-1mm}
where $Head$ is a normalized atom that denotes the predicate that the
assertions apply to, and the $Pre_i$ and $Post_i$ are (DNF) formulas
that refer to the variables of $Head$.
We assume the $Pre_i$ and $Post_i$ to be DNF formulas of \emph{prop}
literals, which specify conditions on the constraint store.
A prop literal $L$ \emph{succeeds trivially} for $\theta$ in program
$P$, denoted $\theta \Rightarrow_P L$, iff $\exists \theta'\in
\answers{(L,\theta)}$ such that $\theta\models\theta'$.

A set of assertions as above states that in any execution state
$\state{Head :: G}{\theta}$ at least one of the $Pre_i$ conditions
should hold, and that, given the $(Pre_i,Post_i)$ pair(s) where
$Pre_i$ holds, then, if $Head$ succeeds, the corresponding $Post_i$
should hold upon success.
More formally, given a predicate represented by a normalized atom
$Head$, and the corresponding set of assertions is $\A = \{A_1 \ldots
A_n\}$, with $A_i = ``\texttt{:- pred } Head \texttt{ : } Pre_i
\texttt{ => }$ $Post_i \texttt{.}$'' such assertions are normalized into
a set of \emph{assertion conditions}
$\{ C_0, C_1, \ldots , C_n\}$,
with:
\vspace*{-4mm}
  \[
    C_i = \left\{
    \begin{array}{ll}
      \callsAsr{Head}{\bigvee _{j = 1}^{n} Pre_j}
    & ~~~~i = 0
    \\
      \successAsr{Head}{Pre_i}{Post_i}
    & ~~~~i = 1..n
    \end{array}
    \right.
  \]

  If there are no assertions associated with $Head$ then
  the corresponding set of conditions is empty.
  The set of assertion conditions for a program is the union of the
  assertion conditions for each of the predicates in the
  program.

The $\callsAsr{Head}{\ldots}$ conditions encode the checks that ensure
that the
calls to the predicate represented by $Head$ are within those
admissible by the set of assertions, and we thus call them the
\emph{calls assertion conditions}.
The $\successAsr{Head_i}{Pre_i}{Post_i}$ conditions encode the checks
for compliance of the successes for particular sets of calls, and we
thus call them the \emph{success assertion conditions}.


We now turn to the operational semantics with assertions, which checks
whether assertion
conditions hold or not while computing the
derivations from a query.  In order to keep track of any violated
assertion conditions, we add labels to the assertion conditions.
Given the atom $L_a$ and the corresponding set of assertion conditions
$\AC$, $\LabAClit{L_a}$ denotes the set of \emph{labeled} assertion
condition instances for $L_a$, where each is of the form
$\asrId{}\#C_a$, such that
$\exists C \in \AC$,
$C = \callsAsr{L}{Pre}$ (or $C = \successAsr{L}{Pre}{Post}$),
$\sigma$ is a renaming s.t. $L = \sigma(L_a)$,
$C_a = \callsAsr{L_a}{\sigma(Pre)}$
(or $C_a = \successAsr{L_a}{\sigma(Pre)}{\sigma(Post)}$),
and $\asrId{}$ is an identifier that is unique for each $C_a$.
We also introduce an extended program state of the form
\exstate{G}{\theta}{\ADeps}, where $\ADeps$ denotes the set of
identifiers for falsified assertion condition instances.
  For the sake of readability, we write labels in
  \emph{negated} form when they appear in the error set.
  We also extend the set of literals with syntactic objects of the
  form $\checkLitLab{L}{\asrId{}}$ where $L$ is a literal and
  $\asrId{}$ is an identifier for an assertion condition instance,
  which we call \emph{check literals}.
Thus, a \emph{literal} is now a constraint, an atom
or a check literal.
We can now recall the notion of \emph{Reductions in Programs with
  Assertions} from~\cite{asrHO-ppdp2014}, which is our starting point:
  a state $S = \exstate{L::G}{\theta}{\ADeps}$, where $L$ is a literal,
  can be \emph{reduced} to a state $S'$, denoted \reductionA{S}{S'},
  as follows:
  \vspace*{-1mm}

  \begin{enumerate}\setlength{\itemsep}{0mm}\setlength{\parskip}{0mm}
  \item
    If $L$ is a constraint or $L = X(t_1,\ldots,t_n)$, then
    $S' = \exstate{G'}{\theta'}{\ADeps}$ where $G'$ and $\theta'$
    are obtained in a same manner as in
    \reduction{\state{L::G}{\theta}}{\state{G'}{\theta'}}

  \item
    If $L$ is an atom and $\exists (L \mbox{\tt :-} B) \in \defn{L}$,
    then $S' = \exstate{B :: G'}{\theta}{\ADeps'}$ where:
    \vspace*{-2mm}
      \[
        \ADeps' = \left\{
        \begin{array}{lr}
          \ADeps \cup \{\negAsrId{}\}
        &
          ~\text{if } \exists \;
          \labCallsAsr{}{L}{Pre} \in \LabAClit{L} \text{ s.t. }
          \theta \not\Rightarrow_P Pre
        \\
          \ADeps
        &
          \text{otherwise}
        \end{array}
        \right.
     \]
     and
     $G' = \checkLitLab{L}{\asrId{1}} :: \ldots :: \checkLitLab{L}{\asrId{n}} :: G$
     such that \\
     $\labSuccessAsr{i}{L}{Pre_i}{Post_i} \in \LabAClit{L} \;\land\;
      \theta \Rightarrow_P Pre_i$.

  \item
    If $L$ is a check literal $\checkLitLab{L'}{\asrId{}}$, then
    $S' = \exstate{G}{\theta}{\ADeps'}$ where
    \vspace*{-2mm}
    \[
      \ADeps' = \left\{
      \begin{array}{lr}
        \ADeps \cup \{\negAsrId{}\}
      &
        ~~\textrm{if}~ \labSuccessAsr{}{L'}{\_}{Post} \in \LabAClit{L'}
        \wedge \theta \not\Rightarrow_P Post
      \\
        \ADeps
      &
        \textrm{otherwise}
      \end{array}
      \right.
    \]

  \end{enumerate}
  \vspace*{-3mm}


\section{Run-time Checking with Caching}
\label{sec:rtc-algo}

The standard operational semantics with run-time checking revisited in
the previous section has the same potential problems as other approaches
which perform exhaustive tests: it can be prohibitively expensive, both
in terms of time and memory overhead.
For example, checking that the first argument of the \texttt{length/2}
predicate is a list at each recursive step turns the standard $O(n)$
algorithm into $O(n^2)$.

As mentioned in the introduction, our objective is to develop an
effective solution to this problem based on memoizing property checks.
An observation that works in our favor is that many of the properties
of interest in the checking process (such as, e.g., regtype
instantiation checking) are monotonic. That is, we will concentrate on
properties such that, for all property checks $L$ if
$\reductionStar{\state{\_}{\theta}}{\state{\_}{\theta'}}$ and $\theta
\Rightarrow_P L$ then $\theta' \Rightarrow_P L$. In this context it
clearly seems attractive to keep $L$ in the store
so that it does not need to be recomputed.
However, memoizing every checked property may also have prohibitive
costs in terms of memory overhead. A worst-case scenario would multiply
the memory needs by the number of call patterns to properties, which can
be large in realistic programs. In addition, looking for stored results
in the store obviously also has a cost that must be taken into account.

\paragraph{Operational Semantics with Caching.}
We base our approach on
an operational semantics which modifies the run-time checking to
maintain and use a \emph{cache store}. The \emph{cache store} $\cache$
is a special constraint store which temporarily holds results from the
evaluation of \emph{prop} literals w.r.t.\ the standard constraint
store $\theta$.
We introduce an extended program state of the form
\exstateM{G}{\theta}{\cache}{\ADeps} and
a \emph{cached} version of ``\emph{succeeds trivially}'': given a
prop literal $L$, it succeeds trivially for $\theta$ and $\cache$ in
program $P$, denoted $\succTrivCache{\theta}{L}$, iff either $L \in
\cache$ or $\succTriv{\theta}{L}$.
Also, the cache store is updated based on the results of the prop
checks, formalized in the following definitions:

\begin{definition}[Updates on the Cache Store]
  Let us consider a DNF formula $Props =
  \bigvee_{i=1}^{n}(\bigwedge_{j=0}^{m(i)}L_{ij})$, where each
  $L_{ij}$ is a prop literal. By $\textsf{lits}(Props) =
  \{L_{ij}|i\in[1:n],j\in[0:m(i)]\}$ we denote the set of all
  literals which appear in $Props$.
  The \emph{cache update} operation is defined as a function $\upd{Props}$ such that:\\
  \centerline{$
    \upd{Props} \subseteq \cache \cup
      \{ L |
          (\succTriv{\theta}{L})
          \wedge
          (L \not\in \cache)
          \wedge
          (L \in \textsf{lits}(Props)\}
  $}
\end{definition}
\vspace*{-1mm}

Note that a precise definition of cache update is left open in this
semantics. Contrary to $\theta$, updates to the cache store $\cache$
are not monotonic since we allow the cache to ``forget'' information
as it fills up, i.e., we assume from the start that $\cache$ is of
limited capacity. However, that information can always be recovered
via recomputation of property checks.
In practice the exact cache behavior depends on parts of the low-level
abstract machine state that are not available at this abstraction
level. It will be described in detail in later sections.

\begin{definition}[Reductions with Assertions and Cache Store]
  \label{def:asr-cache-reductions}
  A state $S = \exstateM{L::G}{\theta}{\cache}{\ADeps}$, where $L$ is a
  literal, can be \emph{reduced} to a state $S'$, denoted
  \reductionA{S}{S'}, as follows:
  \vspace*{-1mm}

  \begin{enumerate}\setlength{\itemsep}{0mm}\setlength{\parskip}{0mm}
  \item
    If $L$ is a constraint or $L = X(t_1,\ldots,t_n)$, then
    $S' = \exstateM{G'}{\theta'}{\cache}{\ADeps}$ where $G'$ and $\theta'$
    are obtained in a same manner as in
    \reduction{\state{L::G}{\theta}}{\state{G'}{\theta'}}

  \item
    If $L$ is an atom and $\exists (L \mbox{\tt :-} B) \in \defn{L}$,
    then \\
    $S' = \exstateM{B :: G'}{\theta}{\cache'}{\ADeps'}$ where:
    \vspace*{-1mm}
      \[
        \begin{array}{rl}
          \ADeps' =
        & \left\{
          \begin{array}{lr}
            \{\negAsrId{}\} \cup \ADeps
          &
            ~\text{if } \exists \;
            \labCallsAsr{}{L}{Pre} \in \LabAClit{L} \text{ s.t. }
            \notSuccTrivCache{\theta}{Pre}
          \\
            \ADeps
          &
            \text{otherwise}
          \end{array}
          \right.
       \vspace*{-1mm}
       \end{array}
     \]

     $\cache' = \upd{Pre}$
     and $G' = \checkLitLab{L}{\asrId{1}} :: \ldots :: \checkLitLab{L}{\asrId{n}} :: G$
     such that \\
     $\labSuccessAsr{i}{L}{Pre_i}{Post_i} \in \LabAClit{L} \;\land\;
      \succTrivCache{\theta}{Pre_i}$.

  \item
    If $L$ is a check literal $\checkLitLab{L'}{\asrId{}}$, then
    $S' = \exstateM{G}{\theta}{\cache'}{\ADeps'}$ where
    \vspace*{-2mm}
    \[
      \begin{array}{rl}
        \ADeps' =
      & \left\{
          \begin{array}{lr}
            \{\negAsrId{}\} \cup \ADeps
          &
            ~~\textrm{if}~ \labSuccessAsr{}{L'}{\_}{Post} \in \LabAClit{L'}
            \wedge \notSuccTrivCache{\theta}{Post}
          \\
            \ADeps
          &
            \textrm{otherwise}
          \end{array}
        \right.
     \end{array}
     \vspace*{-1mm}
   \]
     and $\cache' = \upd{Post}$.
  \vspace*{-1mm}
  \end{enumerate}
\end{definition}


\section{Implementation of Run-time Checking with Caching} 
\label{sec:rtc-impl}

We use the traditional definitional
transformation~\cite{assrt-theoret-framework-lopstr99} as a basis of
our implementation of the operational semantics with cached
checks. This consists of a program transformation that introduces
\emph{wrapper} predicates that check calls and success assertion
conditions while running on a standard (C)LP system. However, we
propose a novel transformation that, in contrast to previous
approaches, groups all assertion conditions for the same predicate
together to produce optimized checks.

Given a program $\P$, for every predicate $p$ the transformation
replaces all clauses $p(\bar{x}) \leftarrow body$ by $p'(\bar{x})
\leftarrow body$, where $p'$ is a new predicate symbol, and inserts
the wrapper clauses given by $\textsf{wrap}(p(\bar{x}),p')$.
The wrapper generator is defined as follows:
\vspace*{-1mm}
\[
  \textsf{wrap}(p(\bar{x}),p') =
    \left\{
    \begin{array}{l}
    \clause{p(\bar{x})}{p_C(\bar{x},\bar{r}),p'(\bar{x}),p_S(\bar{x},\bar{r})}.\\
    \clause{p_C(\bar{x},\bar{r})}{\mathit{ChecksC}}.\\
    \clause{p_S(\bar{x},\bar{r})}{\mathit{ChecksS}}.
    \end{array}
    \right\}
\]
\noindent where $\mathit{ChecksC}$ and $\mathit{ChecksS}$ are the
optimized compilation of pre- and postconditions
$\bigvee_{i=1}^{n}Pre_i$ and $\bigwedge_{i=1}^{n}(Pre_i \rightarrow
Post_i)$ respectively, for
$\labCallsAsr{0}{p(\bar{x})}{\bigvee_{i=1}^{n}Pre_i},$
$\labSuccessAsr{i}{p(\bar{x})}{Pre_i}{Post_i} \in
\LabAClit{p(\bar{x})}$;
and the additional \emph{status} variables $\bar{r}$ are used to
communicate the results of each $Pre_i$ evaluation to the
corresponding $(Pre_i \rightarrow Post_i)$ check.
This way, without any modifications to the literals calling $p$ in the
bodies of clauses in $\P$ (and in any other modules that contain calls
to $p$), after the transformation run-time checks will be performed
for all these calls to $p$ since $p$ (now $p'$) will be accessed via
the wrapper predicate.

The compilation of checks for assertion conditions emits a series of
calls to a \linebreak \kbd{reify\_check(P,R)} predicate, which accepts
as the first argument a property and unifies its second argument with
\kbd{1} or \kbd{0}, depending on whether the property check succeeded or
not.
The results of those reified checks are then combined and evaluated as
boolean algebra expressions using bitwise operations and the
Prolog \texttt{is/2} predicate.
That is, the logical operators $(A \vee B)$, $(A \wedge B)$, and $(A
\rightarrow B)$ used in encoding assertion conditions are replaced by
their bitwise logic counterparts \kbd{R is A \char`\\/ B}, \kbd{R is A /\char`\\
  ~B}, \kbd{R is (A \# 1) \char`\\/ B}, respectively.

The purpose of reification and this compilation scheme is to make it
possible to optimize the logic formulae containing properties that
result from the combination of several \kbd{pred} assertions (i.e., the
assertion conditions).
The optimization consists in reusing the reified status $R$ when
possible, which happens in two ways. First, the \emph{prop} literals
which appear in $Pre$ or $Post$ formulas are only checked once (via
\kbd{reify\_check/2}) and then their reified status $R$ is reused when
needed. Second, the reified status of each $Pre$ conjunction is reused
both in $\mathit{ChecksC}$ and $\mathit{ChecksS}$.

In practice the \textsf{wrap}($p(\bar{x})$,$p'$) clause
generator shares the minimum number of status variables and omits
\emph{trivial} assertion conditions, i.e., those with \texttt{true}
conditions in one of their parts. For instance, excluding
$p_S(\bar{x},\bar{r})$ preserves low-level optimizations such as last
call optimization.%
\footnote{Even though in this work the $p_C(\bar{x},\bar{r})$ and
  $p_S(\bar{x},\bar{r})$ predicates follow the usual bytecode-based
  compilation path, note that they have a concrete structure that is
  amenable to further optimizations (like specialized WAM-level
  instructions or a dedicated interpreter).}

\begin{example}[Program transformation]
\label{ex:asr-transform}
Consider the following annotated program:
\vspace*{-2mm}
\begin{small}
\begin{verbatim}
:- pred p(X,Y) : (int(X) , var(Y)) => (int(X), int(Y)). % A1
:- pred p(X,Y) : (int(X) , var(Y)) => (int(X), atm(Y)). % A2
:- pred p(X,Y) : (atm(X) , var(Y)) => (atm(X), atm(Y)). % A3

p(1,42).  p(2,gamma).  p(a,alpha).
\end{verbatim}
\end{small}
From the set of assertions $\{A1, A2, A3\}$ the following assertion
conditions are constructed:
\vspace*{-4mm}
\begin{small}
  \[
  \begin{array}{rl}
    C_0 =
  & \!\!\!\callsAsr{p(X,Y)}%
                   {(int(X) \wedge var(Y)) \vee ((atm(X) \wedge var(Y)))}
  \\
    C_1 =
  &  \!\!\!\successAsr{p(X,Y)}{(int(X) \wedge var(Y))}%
                              {(int(X) \wedge int(Y))}
  \\
    C_2 =
  &  \!\!\!\successAsr{p(X,Y)}{(int(X) \wedge var(Y))}%
                              {(int(X) \wedge atm(Y))}
  \\
    C_3 =
  &  \!\!\!\successAsr{p(X,Y)}{(atm(X) \wedge var(Y))}%
                              {(atm(X) \wedge atm(Y))}

  \end{array}
  \]
\end{small}
The resulting optimized program transformation is:\\ [1mm]
\begin{tabular}{l|l}
\begin{minipage}{0.45\textwidth}
\begin{small}
\begin{verbatim}
p(X,Y) :-
        p_c(X,Y,R3,R4),
        p'(X,Y),
        p_s(X,Y,R3,R4).

p_c(X,Y,R3,R4) :-
        reify_check(atm(X),R0),
        reify_check(int(X),R1),
        reify_check(var(Y),R2),
        R3 is R1/\R2,
        R4 is R0/\R2,
        Rc is R3\/R4,
        error_if_false(Rc).
\end{verbatim}
\end{small}
\end{minipage}
&
\begin{minipage}{0.45\textwidth}
\begin{small}
\begin{verbatim}
p_s(X,Y,R3,R4) :-
        reify_check(atm(X),R5),
        reify_check(int(X),R6),
        reify_check(atm(Y),R7),
        reify_check(int(Y),R8),
        Rs is (R3#1\/(R6/\R8))
           /\ (R3#1\/(R6/\R7))
           /\ (R4#1\/(R5/\R7)),
        error_if_false(Rs).

p'(1,42).  p'(2,gamma).  p'(a,alpha).
\end{verbatim}
\end{small}
\end{minipage}
\\
\end{tabular}
\ \\
Please note that \kbd{A1} and \kbd{A2} have identical preconditions, and
this is reflected in having only one property combination, \kbd{R3},
for both of them. The same works for individual properties:
in $C_0$ literal $int(X)$ appears twice, literal $var(Y)$
three times, but all such occurrences correspond to only one check in
the code respectively.
\end{example}

The error-reporting predicates \kbd{error\_if\_false/1} in the
instrumented code implement the $\ADeps$ update in the operational
semantics. These predicates abstract away the details of whether
errors produce exceptions, are reported to the user, or are simply
recorded.

The cache itself is accessed fundamentally within the
\kbd{reify\_check/2} predicate. Although the concrete details for a
particular use case (and a corresponding set of experiments) will be
described later, we discuss the main issues and trade-offs involved in
cache implementation in this context.  First, although the cache will
in general be software-defined and dynamically allocated, in any case
the aim is to keep it small with a bounded limit (typically a fraction
of the stacks), so that it does not have a significant impact on the
memory consumption of the program.

Also, in order to ensure efficient lookups and insertions of the cache
elements, it may be advantageous not to store the property calls
literally but rather their memory representation. This means however
that, e.g., for \emph{structure-copying} term representation, a property
may appear more than once in the cache for the same term if its
representation appears several times in memory.

Furthermore, insertion and removal (\emph{eviction}) of entries can be
optimized using heuristics based on the cost of checks (e.g., not
caching simple checks like \texttt{integer/1}), the entry index number
(such as direct-mapped), the history of entry accesses (such as LRU or
least-recently used), or caching \emph{contexts} (such as caching
depth limits during term traversal in regular type checks).

Finally, failure and some of the stack maintenance operations such as
reallocations for stack overflows, garbage collection, or backtracking
need updates on the cache entries (due to invalidation or pointer
reallocation). Whether it is more optimal to evict some or all
entries, or update them is a nontrivial decision that defines another
dimension in heuristics.


\section{Application to Regular Type Checking}
\label{sec:app-regtypes}

As concrete properties to be used in our experiments we select a simple
yet useful subset of the properties than can be used in assertions: the
regular types~\cite{Dart-Zobel} often used in (C)LP systems.
Regular types are properties whose definitions are \emph{regular
  programs}, defined by a set of clauses, each of the form:
``$p(x, v_1, \ldots, v_n) ~\mbox{\tt :-}~ B_1, \ldots, B_k$''
\noindent where $x$ is a linear term (whose variables, which are
called \emph{term variables}, are unique); the terms $x$ of different
clauses do not unify; $v_1$, $\ldots$, $v_n$ are unique variables,
which are called \emph{parametric variables}; and each $B_i$ is either
$t(z)$ (where $z$ is one of the \emph{term variables} and $t$ is a
\emph{regular type expression}) or $q(y, t_1, \ldots, t_m)$ (where
$q/(m+1)$ is a \emph{regular type}, $t_1, \ldots, t_m$ are
\emph{regular type expressions}, and $y$ is a \emph{term variable}).
A \emph{regular type expression} is either a parametric variable or a
parametric type functor applied to some of the parametric variables.
A parametric type functor is a regular type, defined by a regular
program.

\paragraph{Instantiation checks.}
Intuitively, a prop literal $L$ succeeds trivially if $L$ succeeds for
$\theta$ without adding new ``relevant'' constraints to
$\theta$~\cite{prog-glob-an,assert-lang-disciplbook-short}.%
\footnote{Note that checks are performed via entailment checks w.r.t.\
  primitive (Herbrand) constraints. That means that $term(X)$ (which
  is always true) and $ground(X)$ (denoting all possible ground
  terms), despite having the same minimal Herbrand models as
  predicates, do not have the same s-model and are not interchangeable
  as regtype instantiation checks.}
A standard technique to check membership on regular types is based on
\emph{tree automata}. In particular, the regular types defined above are
recognizable by top-down deterministic automata.

This also includes parametric regtypes, provided their parameters are
instantiated with concrete types during checking, since then they can be
reduced to non-parametric regtypes.

Let us recall some basics on deterministic tree automata, as they will
be the basis of our regtype checking algorithm.
A tree automaton is a tuple ${\cal A} = \langle \Sigma, Q, \Delta, Q_f
\rangle$ where $\Sigma$, $Q$, $\Delta$, $Q_f$ are finite sets such
that:
$\Sigma$ is a signature,
$Q$ is a finite set of states,
$\Delta$ is the set of transitions of the form
$f(q_1,\ldots,q_n)\rightarrow q$ where $f \in \Sigma$,
$q,q_1,\ldots,q_n \in Q$ with $n$ being the arity of $f$,
and $Q_f \subseteq Q$ is the set of final states.
The automaton is \emph{top-down deterministic} if $|Q_f|=1$ and for
all $f \in \Sigma$ and all $q \in Q$ there exists at most one sequence
$q_1, \ldots, q_n$ such that $f(q_1,\ldots,q_n) \rightarrow q \in
\Delta$.

Translation of regular types (or instances of parametric regular
types for particular types) from Prolog clauses into 
deterministic top-down tree automata rules is straightforward. This
representation is suitable for low-level encoding (e.g., using
integers for $q_i$ states and a map between each $q_i$ state and its
definition).

\begin{example}
The following \texttt{bintree/2} regular type describes a binary tree
of elements of type \texttt{T}. The corresponding translation into
tree automata rules for the \texttt{bintree(int)} instance with $Q_f =
\{q_b\}$ is shown to its right. \\
\vspace*{1mm}
\begin{tabular}{lr}
\begin{minipage}{0.45\linewidth}
\begin{small}
\begin{verbatim}
:- regtype bintree/2.
bintree(empty,T).
bintree(tree(LC,X,RC),T) :-
    bintree(LC,T),T(X),bintree(RC,T).
\end{verbatim}
\end{small}
\end{minipage}
&
$\begin{array}{rrcl}
\Delta = \{ & empty & \rightarrow & q_b \\
& tree(q_b, q_{int}, q_b) & \rightarrow & q_b ~\}
\end{array}$
  \end{tabular}
\vspace*{-4mm}
\end{example}

\paragraph{Algorithm for Checking Regular Types with Caches.}

We describe the \textsc{RegCheck} algorithm for regtype checking using
caches in Algorithm~\ref{alg:regcheck}. The \kbd{reify\_check/2}
predicate acts as the interface between \textsc{RegCheck} and the
runtime checking framework.
\begin{algorithm}[t]
\caption{Check that the type of the term stored at $x$ is $t$, at depth $d$.}
\label{alg:regcheck}
\small
\begin{algorithmic}
\Function{RegCheck}{$x, t, d$}
  \State Find $C \in \textit{Constructors}(t)$ so that $Functor(C) = Functor(x)$,
  \State otherwise \Return False
  \If{$Arity(x)=0$}
    \Comment{Atomic value, not cached}
    \State \Return True 
  \ElsIf{\Call{CacheLookup}{$x,t$}}
    \Comment{Already in cache}
    \State \Return True
  \ElsIf{$\forall i \in [1,Arity(x)] . \Call{RegCheck}{Arg(i,x), Arg(i,C), d+1}$}
    \If{$d < depthLimit$}
      \Comment{Insert in cache}
      \State \Call{CacheInsert}{$x, t$}
    \EndIf
    \State \Return True \Comment{In regtype}
  \Else
    \State \Return False \Comment{Not in regtype}
  \EndIf
\EndFunction
\end{algorithmic}
\vspace*{-1mm}
\end{algorithm}
The algorithm is derived from the standard definition of \emph{run} on
tree automata.
A run of a tree automaton ${\cal A} = \langle \Sigma, Q, \Delta, Q_f
\rangle$ on a tree $x \in T_{\Sigma}$ (terms over $\Sigma$) is a
mapping $\rho$ assigning a state to each occurrence (subterm) of
$f(x_1,\ldots,x_n)$ of $x$ such that:
\vspace*{-3mm}
\[
f(\rho(x_1),\ldots,\rho(x_n)) \rightarrow \rho(f(x_1,\ldots,x_n)) \in \Delta
\]
A term $x$ is recognized by ${\cal A}$ if $\rho(x) \in Q_f$.
For deterministic top-down recognition, the algorithm starts with the
single state in $Q_f$ (which for simplicity, we will use to identify
each regtype and its corresponding automata) and follows the rules
\emph{backwards}.
The tree automata transition rules for a regtype are consulted with
the functions $\textit{Constructors}(t) = \{C | C \rightarrow t \in
\Delta\}$, $Arg(i,u)$ (the i-th argument of a constructor or term
$u$), and $Functor(u)$ (the functor symbol, including arity, of a
constructor or term $u$). Once there is a functor match, the regtypes
of the arguments are checked recursively.
To speed up checks, the cache is consulted
($\textsc{CacheLookup}(x,t)$ searches for $(x,t)$) before performing
costly recursion, and \emph{valid} checks inserted
($\textsc{CacheInsert}(x,t)$ inserts $(x,t)$) if needed (e.g., using
heuristics, explained below).
The cache for storing results of regular type checking is implemented
as a \emph{set} data structure that can efficiently insert and look up
$(x,t)$ pairs, where $x$ is a term address\footnote{Since regtype
  checks are monotonic, this is safe as long as cache entries are
  properly invalidated on backtracking, stack movements, and garbage
  collection. Using addresses is a pragmatic decision to minimize the
  overheads of caching.}
and $t$ a regular type identifier. The specific implementation depends
on the cache heuristics, as described below.

\paragraph{Complexity.} It is easy to show that complexity has
$O(1)$ best case (if $x$ was cached) and $O(n)$ worst case, with $n$
being the number of tree nodes (or term size). In practice, the
caching heuristics can drastically affect performance. For example,
assume a full binary tree of $n$ nodes. Caching all nodes at levels
multiple of $c$ will need $n/(2^{c+1}-1)$ entries, with a constant
cost for the worst case check (at most $2^{c+1}-1$ will be checked,
independently of the size of the term).

\paragraph{Cache Implementation and Heuristics.}
In order to decide what entries are added and what entries are evicted
to make room for new entries on cache misses, we have implemented
several caching heuristics and their corresponding data
structures. Entry eviction is controlled by \emph{replacement
  policies}:
\vspace*{-2mm}
\begin{itemize}
\item Least-recently used (LRU) replacement and fully
  associative. Implemented as a hash table whose entries are nodes of a
  doubly linked list.
  The most recently accessed element is moved to the head and new
  elements are also added to the head. If cache size exceeds the maximal size
  allowed, the cache is pruned.
\item Direct-mapped cache with collision replacement, with a simple
  hash function based on modular arithmetic on the term address. This
  is simpler but less predictable.
\end{itemize}

The insertion of new entries is controlled by the caching
\emph{contexts}, which include the regular type being checked and the
location of the check:
\vspace*{-2mm}
\begin{itemize}
\item We do not cache simple properties (like primitive type tests,
  e.g., \texttt{integer/1}, etc), where caching is more expensive than
  recomputing.
\item We use the check depth level in the cache interface for recursive
  regular types.
  Checks beyond this threshold depth limit are not cached. This
  gives priority to roots of data structures over internal
  subterms which may pollute the cache.
\end{itemize}

\paragraph{Low-level C implementation.}
In our prototype, this algorithm is implemented in C with some
specialized cases (as required for our WAM-based representation of
terms, e.g., to deal with atomic terms, list constructors, etc.).%
\footnote{Even though the algorithm can be easily implemented as a
  deterministic Prolog program, we chose in this work a specialized,
  lower-level implementation that can interact more directly with the
  optimized cache data structures.} 
The regtype definition is encoded as a map between functors (name and
arity) and an array of $q$ states for each argument. For a small number
of functors, the map is implemented as an array. Efficient lookup for
many functors is achieved using hash maps.
Additionally, a number of implicit transition rules exist for
primitive types (any term to $q_{any}$, integers to $q_{int}$, etc.)
that are handled as special cases.


\section{Experimental Results and Evaluation}
\label{sec:eval}

To study the impact of caching on run-time overhead, we have evaluated
the run-time checking framework on a set of 7 benchmarks, for regular
types. We consider benchmarks where we perform a series of element
insertions in a data structure.
Benchmarks \kbd{amqueue}, \kbd{set}, \kbd{B-tree}, and (binary)
\kbd{tree} were adapted from the Ciao libraries; benchmarks \kbd{AVl-tree},
\kbd{RB-tree} and \kbd{heap} were adapted from the YAP libraries. These
benchmarks can be divided into 4 groups:
\vspace*{-2mm}
\begin{enumerate}[(a)]
\item simple list-based data structures: \kbd{amqueue}, \kbd{set};
\item balanced tree-based structures that do not change the structural
properties of their nodes on balancing: \kbd{AVl-tree}, \kbd{heap};
\item balanced tree-based structures that change node properties:
\kbd{B-tree} (changes the number of node children), \kbd{RB-tree}
(changes node color);
\item unbalanced tree structures (\kbd{tree}).
\end{enumerate}

For each run of the benchmark suite the following parameters were
varied: cache replacement policy (LRU, direct mapping), cache size (1 to
256 cells), and check depth threshold (1 to 5, and ``infinite''
threshold for unlimited check depth).
Table~\ref{tbl:bmks-results} summarizes the results of the experiments.
For each combination of the parameters it reports the optimal caching
policy, LRU (L) or direct mapping (D). Also, for each of the benchmarks
it reports an interval within which the worst case check depth varies.

\begin{table}[t]
\begin{center}
  \caption{Benchmarks}
  \begin{tabular}{r|c||rlrl|rlrl|rlrl|rl}
   \cline{1-16} \noalign{\vskip 0.6mm}
    \multicolumn{2}{r||}{\textsf{benchmark}}
    & \multicolumn{2}{c}{\skbd{amqueue}}
    & \multicolumn{2}{c|}{\skbd{set}}
    & \multicolumn{2}{c}{\skbd{AVL-tree}}
    & \multicolumn{2}{c|}{\skbd{heap}}
    & \multicolumn{2}{c}{\skbd{B-tree}}
    & \multicolumn{2}{c|}{\skbd{RB-tree}}
    & \multicolumn{2}{c}{\skbd{tree}}
  \\
    \multicolumn{2}{r||}{\textsf{assertions}}
    & \multicolumn{2}{c}{4}
    & \multicolumn{2}{c|}{4}
    & \multicolumn{2}{c}{8}
    & \multicolumn{2}{c|}{7}
    & \multicolumn{2}{c}{9}
    & \multicolumn{2}{c|}{15}
    & \multicolumn{2}{c}{2}
  \\
    \multicolumn{2}{r||}{\textsf{regtypes}}
    & \multicolumn{2}{c}{1}
    & \multicolumn{2}{c|}{1}
    & \multicolumn{2}{c}{1}
    & \multicolumn{2}{c|}{2}
    & \multicolumn{2}{c}{5}
    & \multicolumn{2}{c|}{2}
    & \multicolumn{2}{c}{1}
  \\ \noalign{\vskip 0.6mm} \cline{1-16} \noalign{\vskip 0.6mm}
    \multicolumn{2}{c||}{\textsf{depth limit}}
    & \textsf{2} & $\infty$
    & 2 & $\infty$
    & 2 & $\infty$
    & 2 & $\infty$
    & 2 & $\infty$
    & 2 & $\infty$
    & 2 & $\infty$
  \\
   \multirow{2}{*}{\textsf{cache}}
    & 256
    & D & D
    & D & D
    & L & L
    & L & L
    & L & L
    & L & D
    & D & L
  \\
    & 128
    & D & D
    & D & D
    & L & L
    & L & L
    & D & D
    & L & D
    & D & D
  \\
  \multirow{2}{*}{\textsf{size}}
    & 64
    & D & D
    & D & D
    & L & L
    & L & L
    & D & D
    & L & D
    & D & D
  \\
    & 32
    & D & D
    & D & D
    & D & D
    & D & D
    & D & D
    & D & D
    & D & D
  \\ \noalign{\vskip 0.6mm} \cline{1-16} \noalign{\vskip 0.6mm}
    \textsf{max}
    & DM
    & \multicolumn{2}{c}{$2$}
    & \multicolumn{2}{c|}{$1$}
    & \multicolumn{2}{c}{$[7:11]$}
    & \multicolumn{2}{c|}{$[5:11]$}
    & \multicolumn{2}{c}{$[13:21]$}
    & \multicolumn{2}{c|}{$[6:21]$}
    & \multicolumn{2}{c}{$[9:20]$}
  \\
    \textsf{depth}
    & LRU
    & \multicolumn{2}{c}{$2$}
    & \multicolumn{2}{c|}{$1$}
    & \multicolumn{2}{c}{$[3:11]$}
    & \multicolumn{2}{c|}{$[1:11]$}
    & \multicolumn{2}{c}{$[4:21]$}
    & \multicolumn{2}{c|}{$[6:20]$}
    & \multicolumn{2}{c}{$[6:20]$}
  \\ \noalign{\vskip 0.6mm}
  \cline{1-16}
  \end{tabular}
  \label{tbl:bmks-results}
\end{center}
\vspace*{-3mm}
\end{table}

\begin{figure}[ht!]
  \centering
  \begin{subfigure}[b]{0.32\textwidth}
    \includegraphics[width=\textwidth]{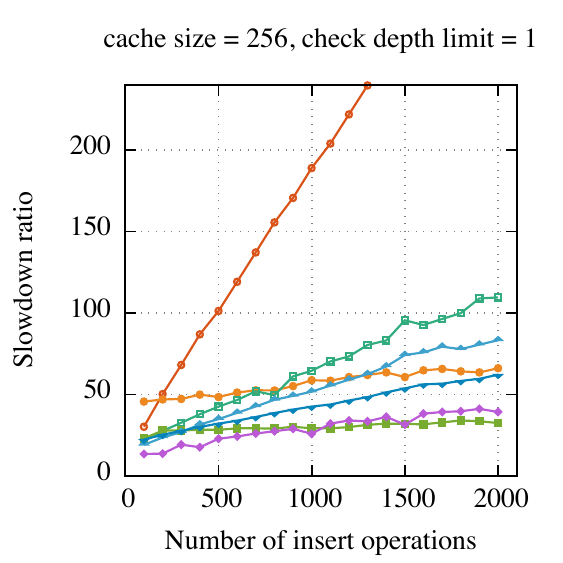}
  \end{subfigure}%
  ~
  \begin{subfigure}[b]{0.32\textwidth}
    \includegraphics[width=\textwidth]{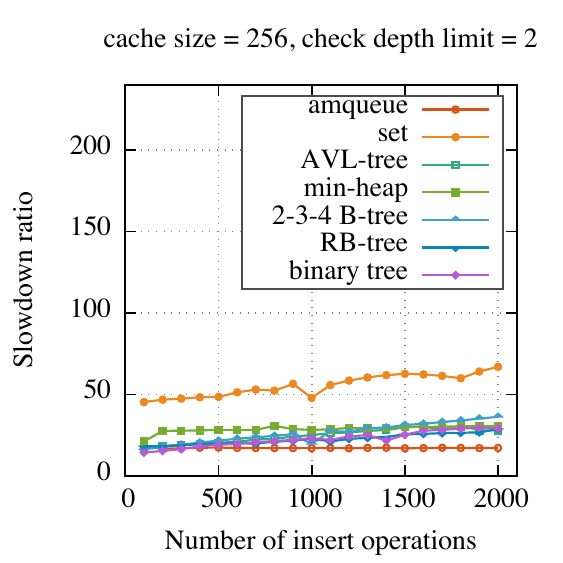}
  \end{subfigure}
  ~
  \begin{subfigure}[b]{0.32\textwidth}
    \includegraphics[width=\textwidth]{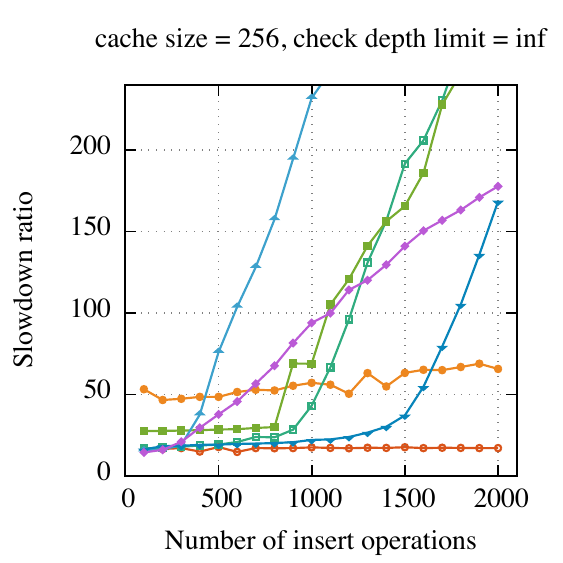}
  \end{subfigure}

  \begin{subfigure}[b]{0.32\textwidth}
    \includegraphics[width=\textwidth]{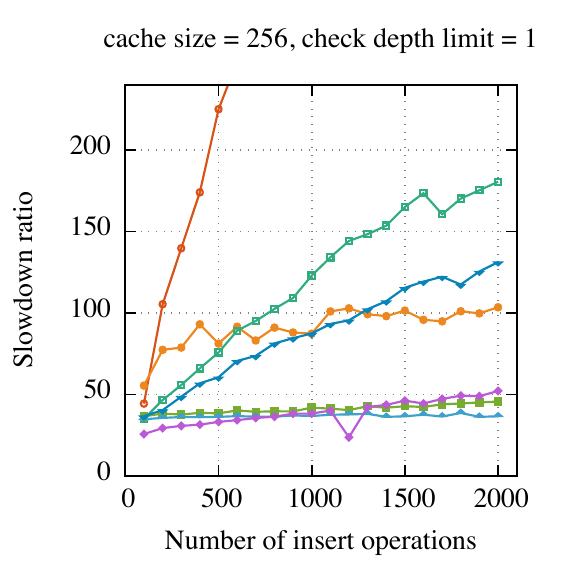}
  \end{subfigure}%
  ~
  \begin{subfigure}[b]{0.32\textwidth}
    \includegraphics[width=\textwidth]{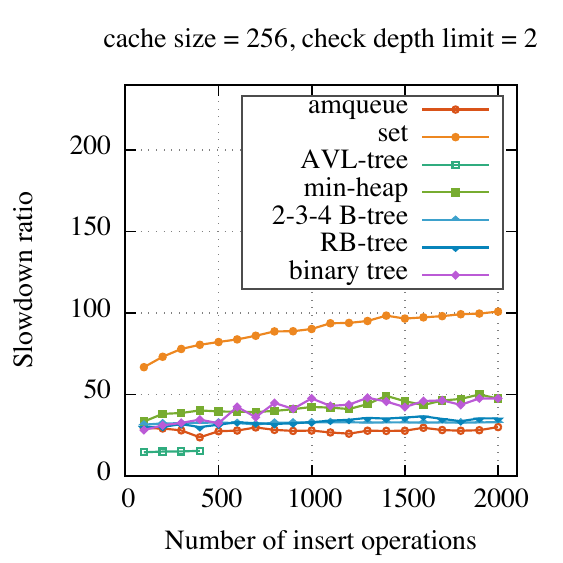}
  \end{subfigure}
  ~ 
  \begin{subfigure}[b]{0.32\textwidth}
    \includegraphics[width=\textwidth]{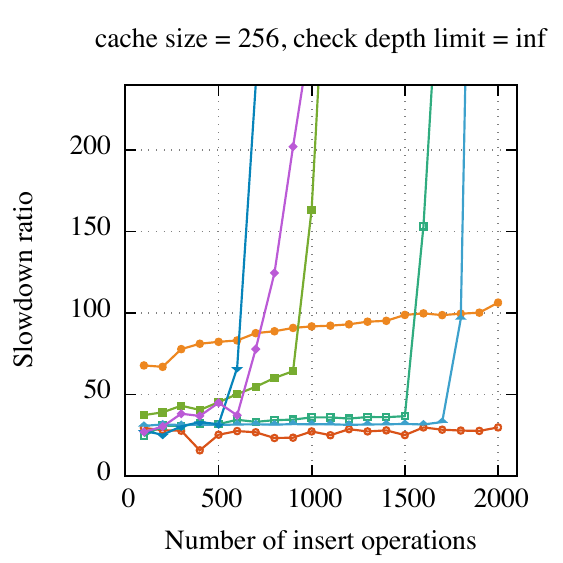}
      \end{subfigure}

  \caption{Run-time check overhead ratios for all benchmarks with check
     depth thresholds of 1, 2, $\infty$, and DM (top row) and LRU (bottom
     row) policies in cache of 256 elements.}
  \label{fig:InfWorse2}
  \vspace*{-3mm}
\end{figure}

The experiments show that the overhead of checks with depth threshold
2 (storing the regtype of the check argument and the regtypes of its
arguments) is smaller than or equal to the one obtained with unlimited
depth limit~(Fig~\ref{fig:InfWorse2}).
A depth limit of 1 does not allow checks to store enough useful
information about terms of most of the data structures (compare the
overhead increase for \skbd{amqueue} with this and bigger limits),
while unlimited checks tend to overwrite this information multiple
times, so that it cannot be reused.
At the same time, for data structures represented by large nested terms
(e.g., nodes of \skbd{B-trees}), deeper limits (3 or 4) for small inputs
seem more beneficial for capturing such term structure.
It can also be observed that the lower cost of element insert/lookup
operations with the DM cache replacement policy results in having
lower total overhead than with the LRU policy.

\begin{figure}[t]
  \centering
  \begin{subfigure}[b]{0.35\textwidth}
    \includegraphics[width=\textwidth]{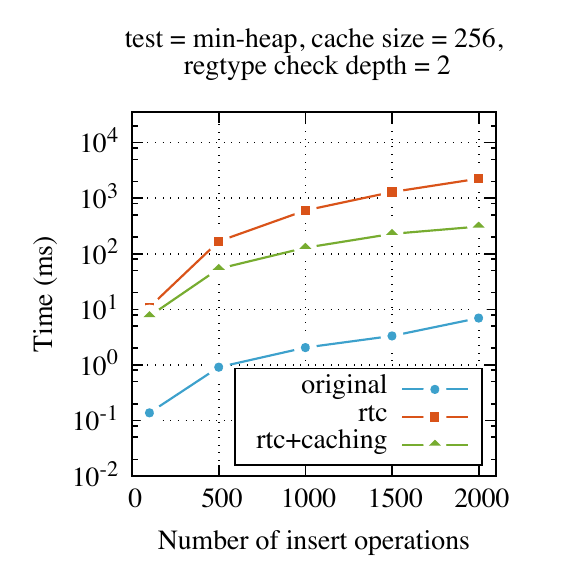}
  \end{subfigure}%
  ~%
  \begin{subfigure}[b]{0.35\textwidth}
    \includegraphics[width=\textwidth]{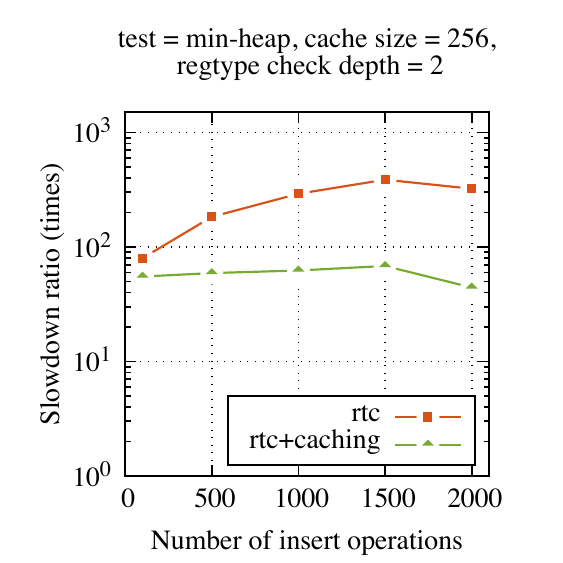}
  \end{subfigure}%
  \caption{Absolute and relative running times of the \skbd{heap}
  benchmark with different rtchecks configurations, LRU caching policy.}
  \label{fig:abs_rel}
\end{figure}

While even with caching the cost of the run-time checks still remains
significant,\footnote{Note that in general run-time checking is a
  technique for which non-trivial overhead can be expected for all but
  the most trivial properties. It can be conceptually associated with
  running the program in the debugger, which typically also introduces
  significant cost.}  caching does reduce overhead by 1-2 orders of
magnitude with respect to the cost of run-time checking without
caching~(Fig.~\ref{fig:abs_rel}).  Also, the slowdown ratio of
programs with run-time checks using caching is almost constant, in
contrast with the linear (or worse) growth in the case where caching
is not used. An important issue that has to be taken into account here
is that most of the benchmarks are rather simple, and that performing
insert operations is much less costly that performing run-time checks
on the arguments of this operation. This explains the observation that
checking overhead is the highest for the \kbd{set}
benchmark~(Fig~\ref{fig:InfWorse2}), while it is one of the simplest
used in the experiments.

\begin{figure}[t]
  \centering
  \begin{subfigure}[b]{0.35\textwidth}
    \includegraphics[width=\textwidth]{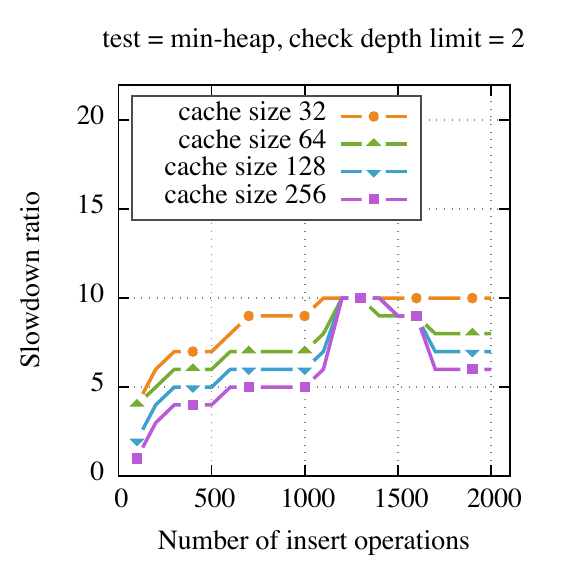}
    \caption{LRU}
  \end{subfigure}
  ~%
  \begin{subfigure}[b]{0.35\textwidth}
    \includegraphics[width=\textwidth]{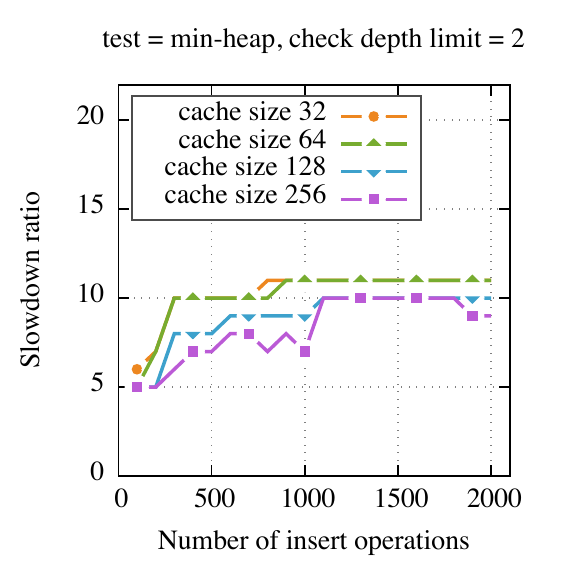}
    \caption{DM}
  \end{subfigure}%
  \caption{Worst case regtype check depth for benchmarks from groups (b)
  and (c), with LRU and DM cache replacement policies respectively.}
  \label{fig:allRT_b_c}
\end{figure}

Another factor that affects the overhead ratio is cache size.
For smaller caches cell rewritings occur more often, and thus the
optimal cache replacement policy in such cases is the one with the
cheapest operations.
For instance, for cache size 32 the optimal policy for all benchmark
groups is DM, while for other cache sizes LRU is in some cases better as
it allows optimizing cell rewritings.
This observation is also confirmed by the maximal check depth in the
worst case, which is almost half on average for the
benchmarks for which LRU is the optimal policy
(Fig~\ref{fig:allRT_b_c}).
In the simple data structures of group (a) the experiments show that it is
beneficial to have cheaper cache operations (like those of caches with
DM caching policy), since such structures do not suffer from cache cell
rewritings as much as more complex structures.
The same observation is still true for group (d), where for some inputs
the binary tree might grow high and regtype checks of leaves will pollute
the cache with results of checks for those inner nodes on the path, that
are not in the cache, overwriting cache entries with regtypes of
previously checked nodes.
The DM policy also happens to show better results for group (c) for a
similar reason. Since data structures in this group change essential
node properties during the tree insertion operation, this in practice
means that sub-terms that represent inner tree nodes are (re-)created
more often. As a result, with the LRU caching policy the cache would
become populated by check results for these recently created nodes,
while the DM caching policy would allow preserving (and reusing) some
of the previously obtained results.
The only group that benefits from LRU is (b), where this policy helps
preserving check results for the tree nodes that are closer to the root
(and are more frequently accessed) and most of the overwrites happen to
cells that store leaves.

More plots are available in the online appendix (Appendix A).


\section{Conclusions and Related Work}
\label{sec:concl}

We have presented an approach to reducing the overhead implied by
run-time checking of properties based on the use of memoization to
cache intermediate results of check evaluation, avoiding repeated
checking of previously verified properties. We have provided an
operational semantics with assertion checks and caching and an
implementation approach, including a more efficient program
transformation than in previous proposals. We have also reported on a
prototype implementation and provided experimental results that
support that using a relatively small cache leads to very significant
decreases in run-time checking overhead.
The idea of using memoization techniques to speed up checks has
attracted some attention recently~\cite{rv2014}.  Their work
(developed independently from ours) is based on adding fields to data
structures to store the properties that have been checked already for
such structures.
In contrast, our approach has the advantage of not requiring any
modifications to data structure representation, or to the checking code,
program, or core run-time system.
Compared to the approaches that reduce checking frequency our proposal
has the advantage of being exhaustive (i.e., all tests are checked at
all points) while still being much more efficient than standard
run-time checking.
Our approach greatly reduces the overhead when tests are being
performed, while allowing the parts for which testing is turned off to
execute at full speed without requiring recompilation.
While presented for concreteness in the context of the Ciao run-time
checking framework, we argue that the approach is general, and the
results should carry over to other programming paradigms.

\vspace*{-1mm}
\paragraph*{\textbf{Acknowledgments:}}
Research supported in part by projects EU FP7 318337 \emph{ENTRA},
Spanish MINECO TIN2012-39391 \emph{StrongSoft},
and Madrid Regional Government
S2013/ICE-2731, N-Greens Software.

\bibliographystyle{acmtrans}

\clearpage


\section*{Supplementary material}

Online appendix for the paper

# Practical Run-time Checking
# via Unobtrusive Property Caching

published in Theory and Practice of Logic Programming

NATALIIA STULOVA[1]    JOSÉ F. MORALES[1]    MANUEL V. HERMENEGILDO[1,2]

[1]*IMDEA Software Institute*
(*e-mail:* {`nataliia.stulova, josef.morales, manuel.hermenegildo`}`@imdea.org`)
[2]*School of Computer Science, Technical University of Madrid (UPM)*
(*e-mail:* `manuel.hermenegildo@upm.es`)

## Appendix A

This appendix includes plots of the run-time checking overhead observed in the set of 7 benchmarks for different cache replacement policies. There are four groups of plots:

- overhead ratio plots, where overhead ratio curves are grouped by cache size and check depth limit (Figures A 1 and A 5);
- overhead ratio plots, where overhead ratio curves are grouped by benchmark and check depth limit (Figures A 2 and A 6);
- maximal regtype check depth reached plots, where check depth curves are grouped by benchmark and cache size (Figures A 3 and A 7);
- absolute and relative benchmark execution time plots for benchmarks without rtchecks, with rtchecks and with both rtchecks and caching (Figures A 4 and A 8).

## A.1 "Least Recently Used" Cache Replacement Policy

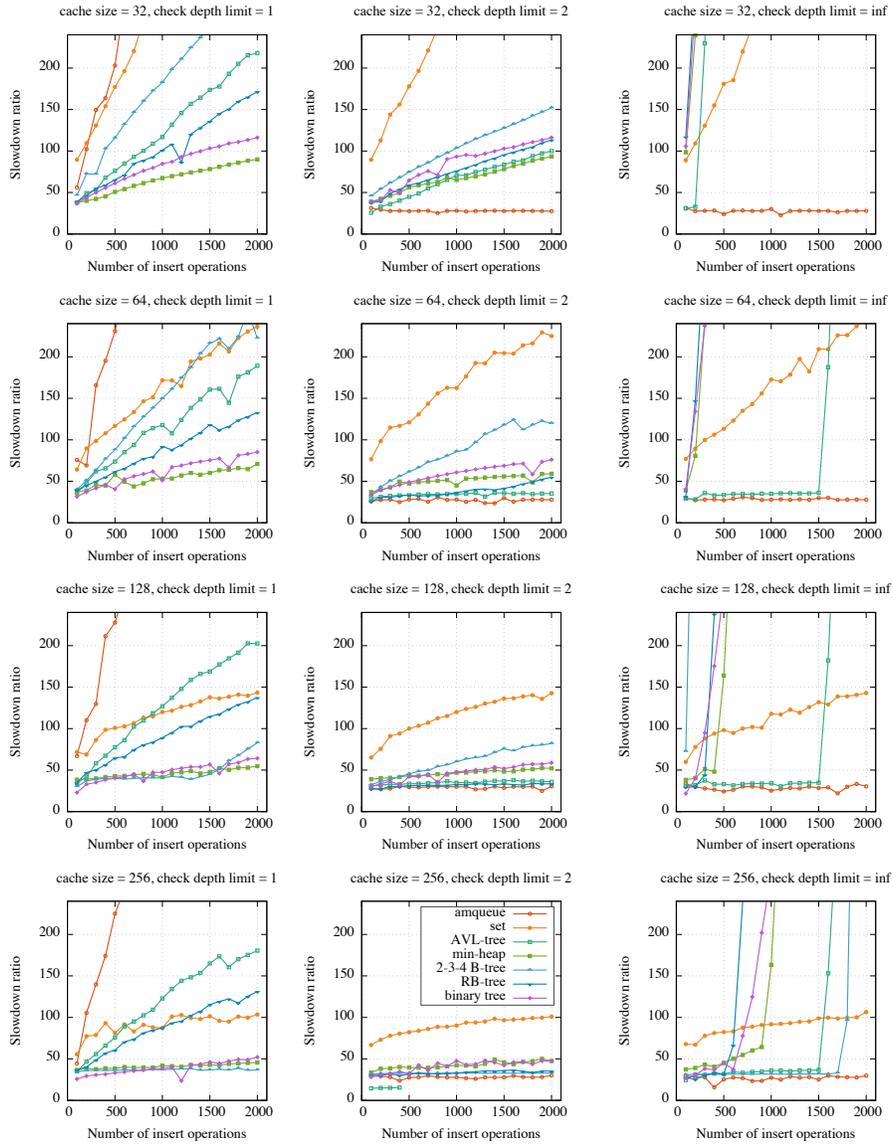

Fig. A 1: Overhead ratios for all benchmarks, check depth limits 1, 2 and $\infty$, LRU caching policy.

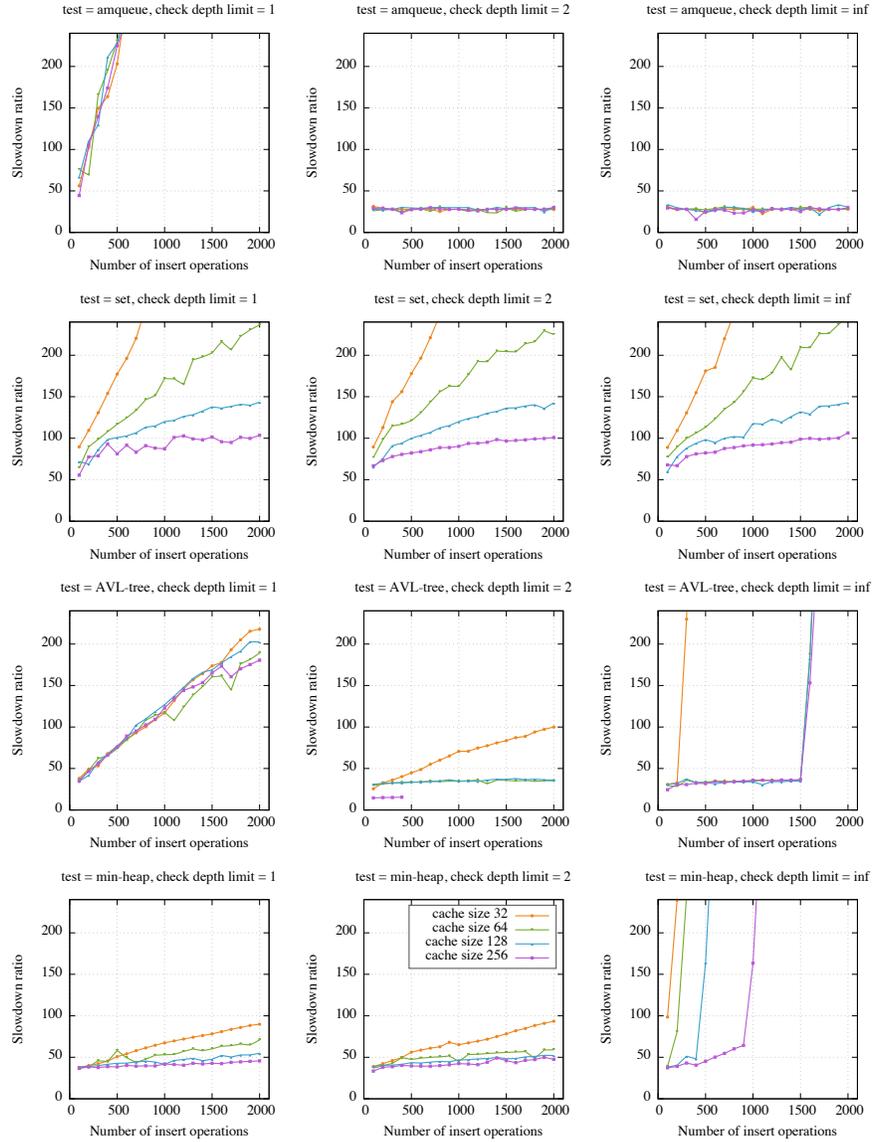

Fig. A 2: Overhead ratios for each benchmark, check depth limits 1, 2 and $\infty$, LRU caching policy.

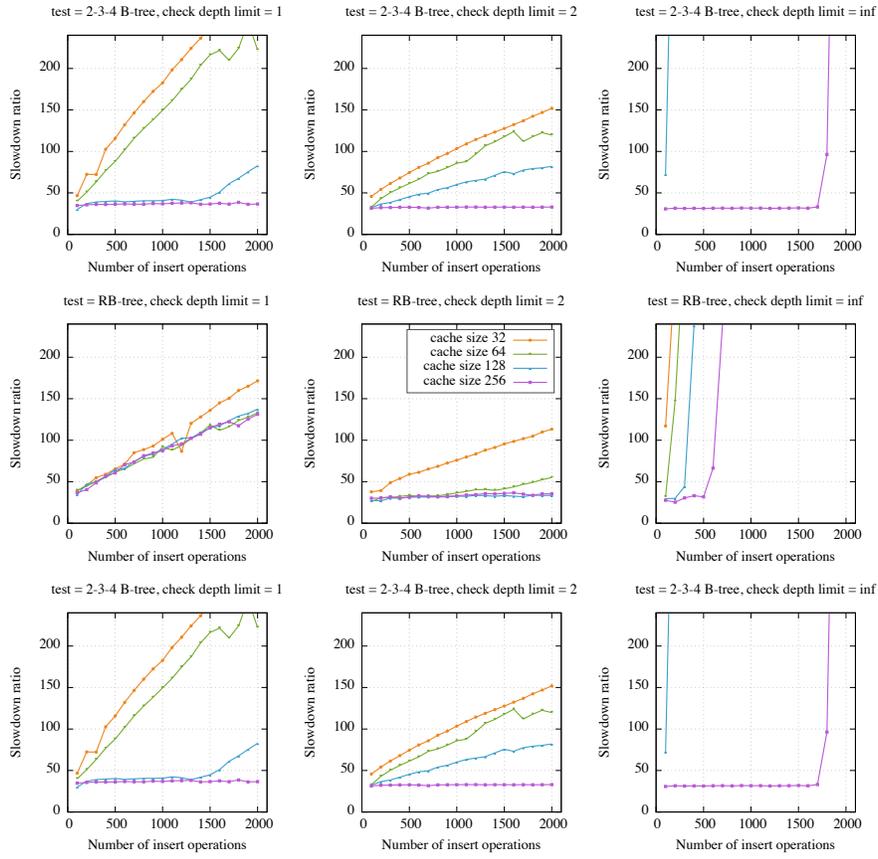

Fig. A 2: Overhead ratios for each benchmark, check depth limits 1, 2 and ∞, LRU caching policy.

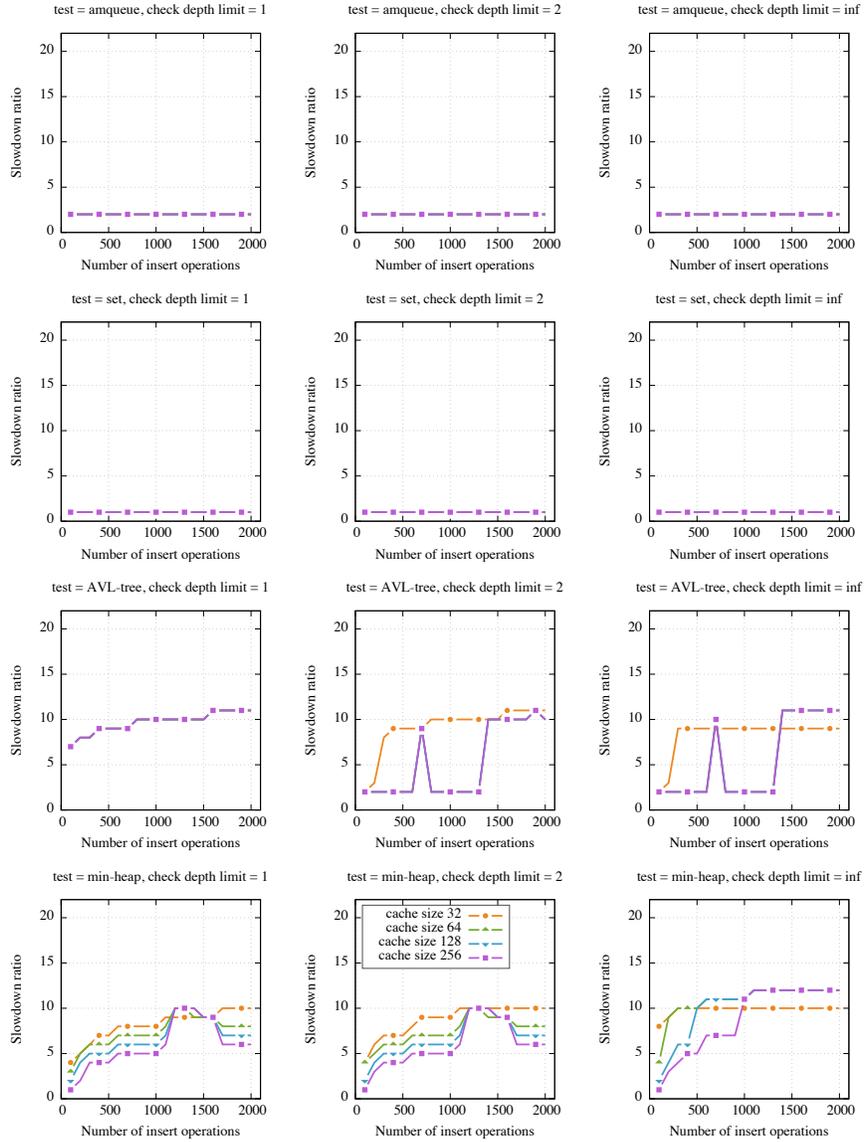

Fig. A 3: Max regtype check depth for each benchmark, check depth limits 1, 2 and $\infty$, LRU caching policy.

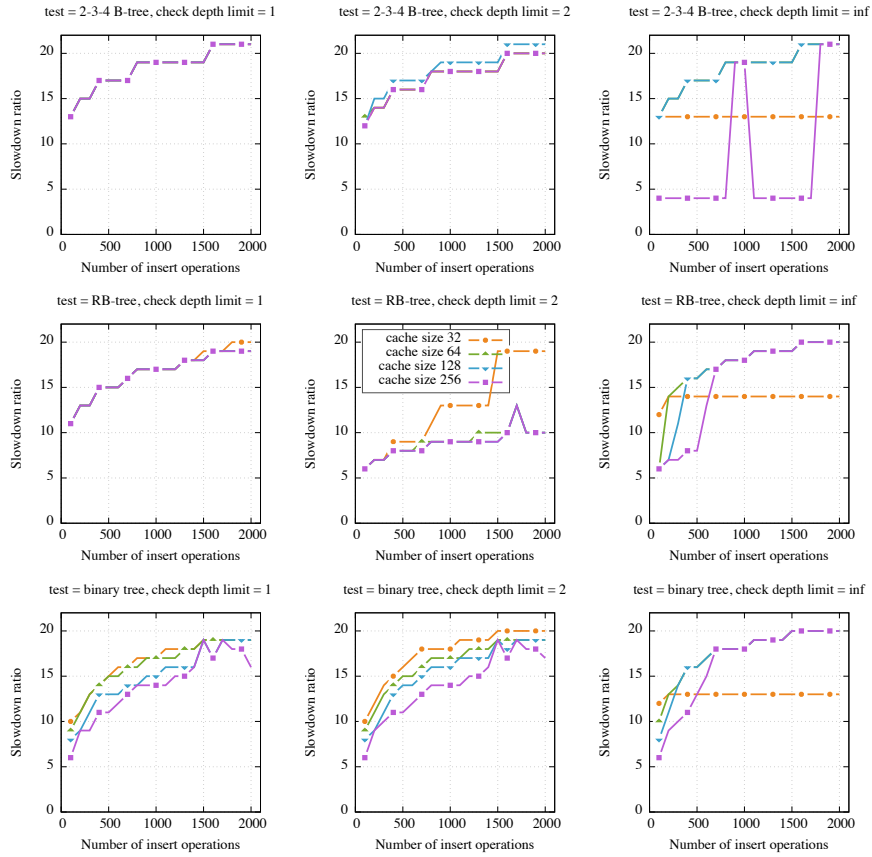

Fig. A 3: Max regtype check depth for each benchmark, check depth limits 1, 2 and ∞, LRU caching policy.

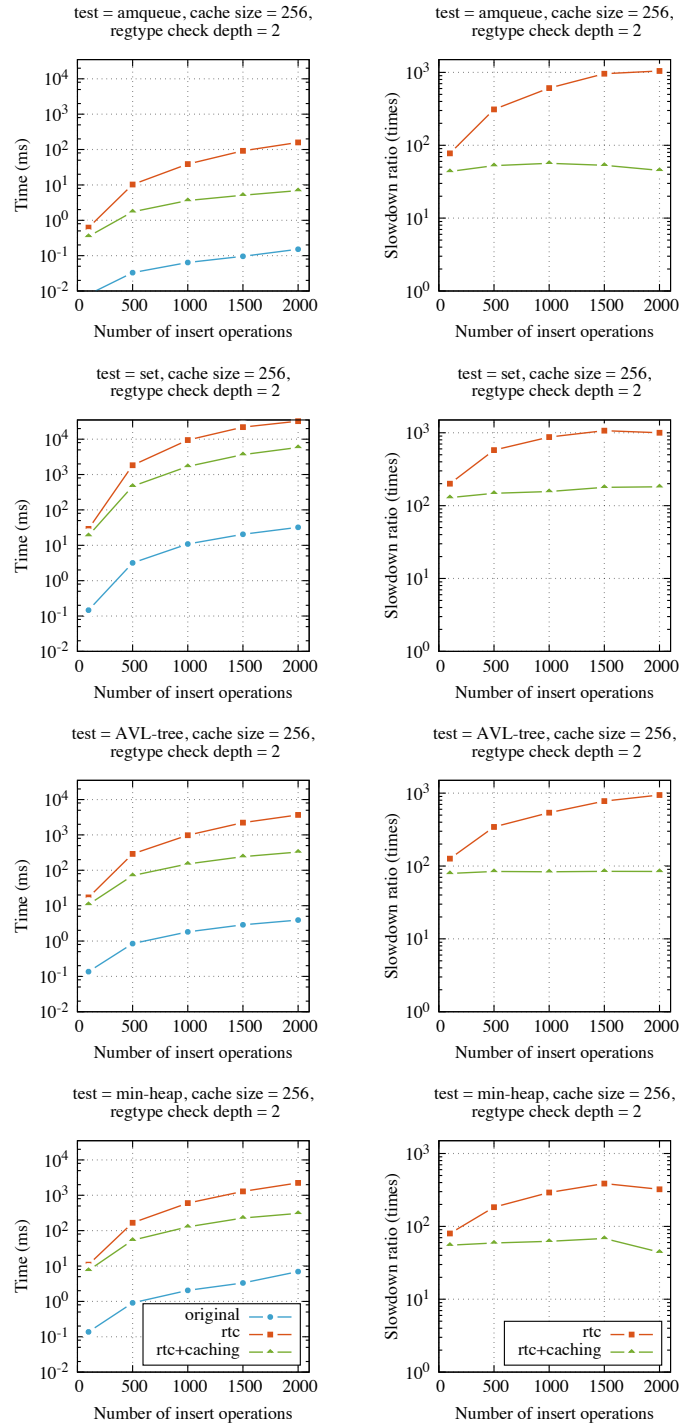

Fig. A 4: Absolute and relative benchmark running times, cache size 256 elements, check depth limit 2, LRU caching policy.

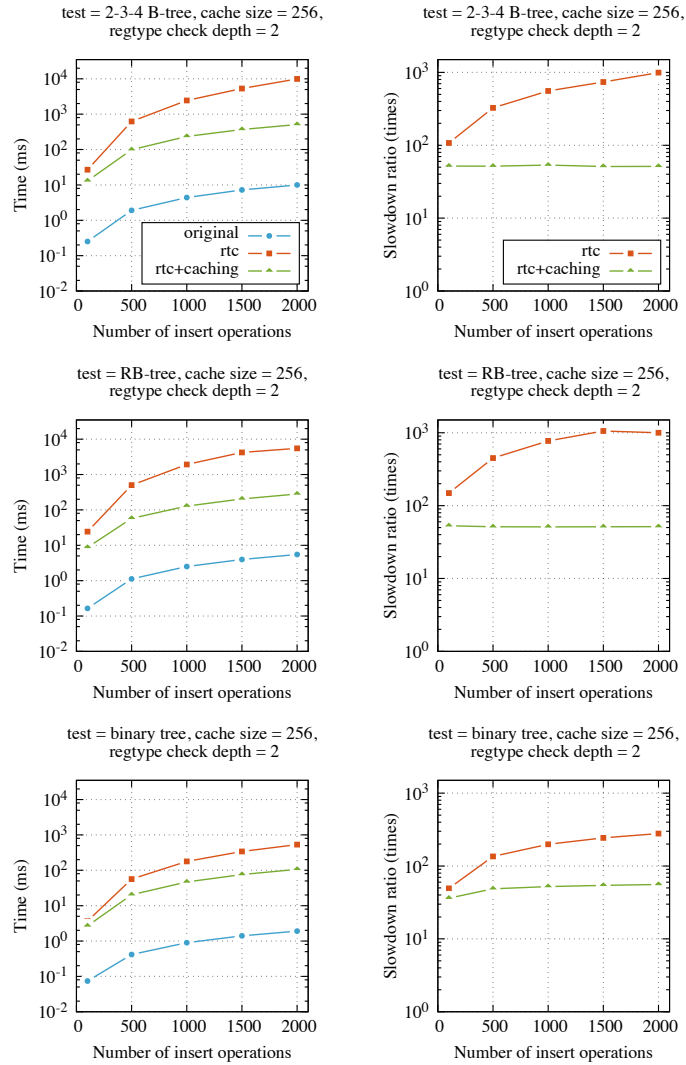

Fig. A 4: Absolute and relative benchmark running times, cache size 256 elements, check depth limit 2, LRU caching policy.

## A.2 Direct Mapping Cache Replacement Policy

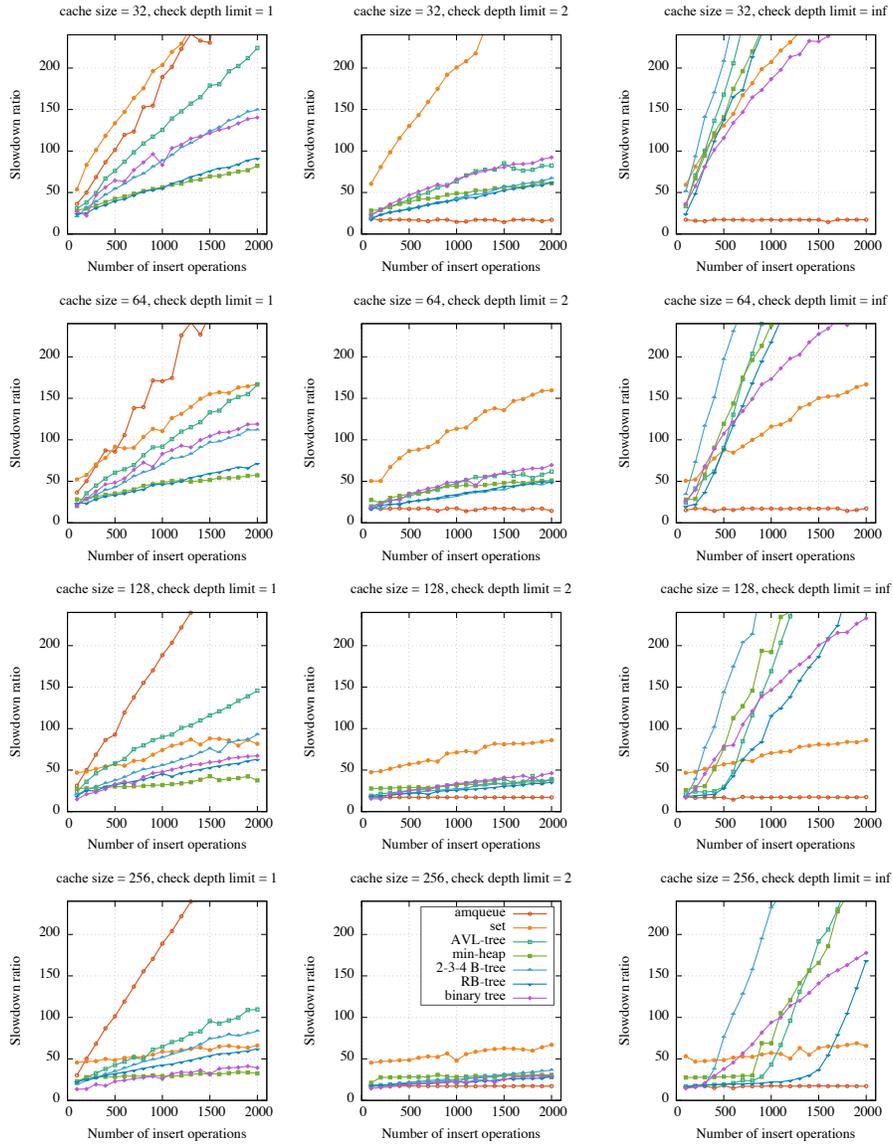

Fig. A 5: Overhead ratios for all benchmarks, check depth limits 1, 2 and $\infty$, DM caching policy.

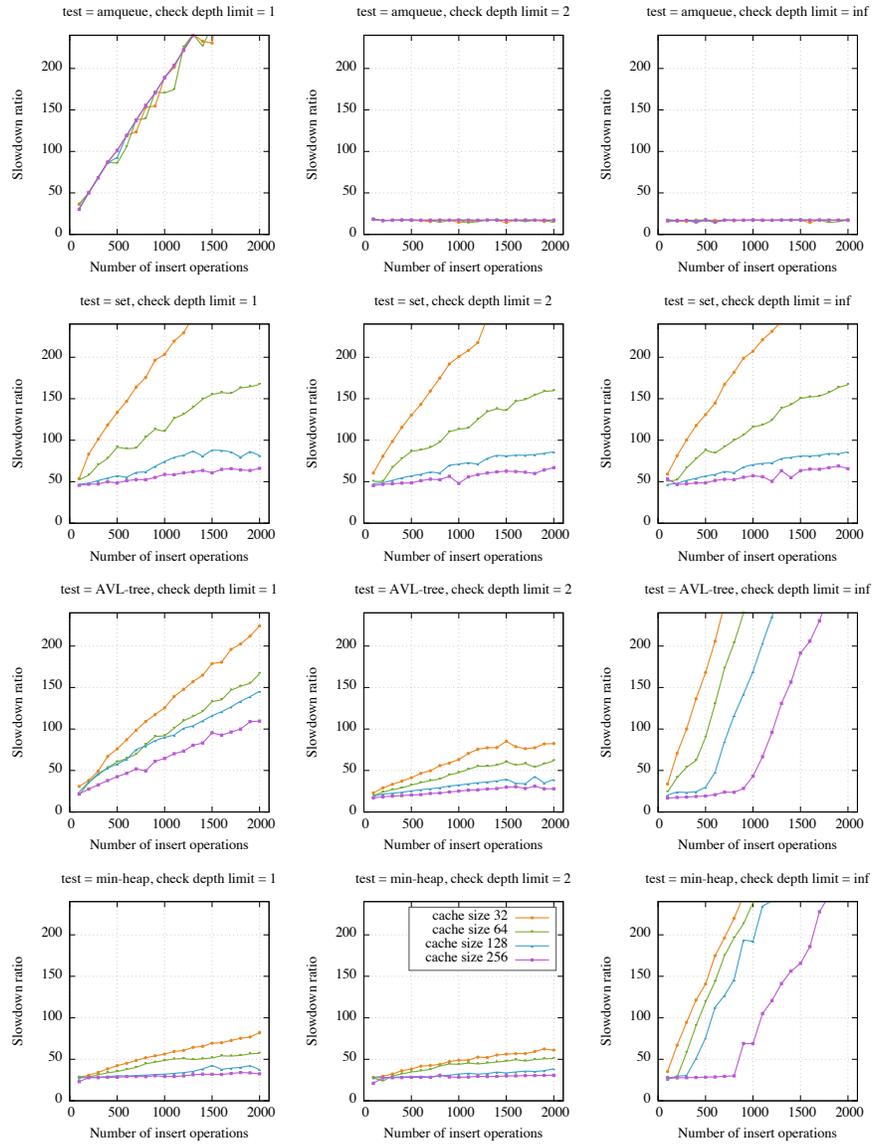

Fig. A 6: Overhead ratios for each benchmark, check depth limits 1, 2 and $\infty$, DM caching policy.

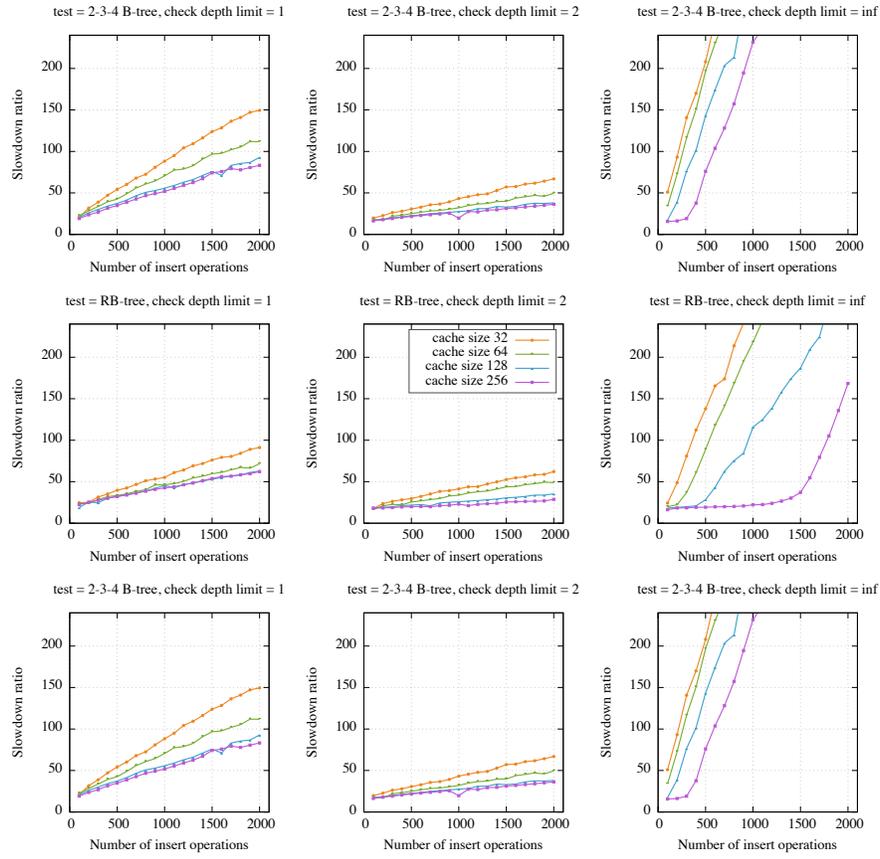

Fig. A 6: Overhead ratios for each benchmark, check depth limits 1, 2 and $\infty$, DM caching policy.

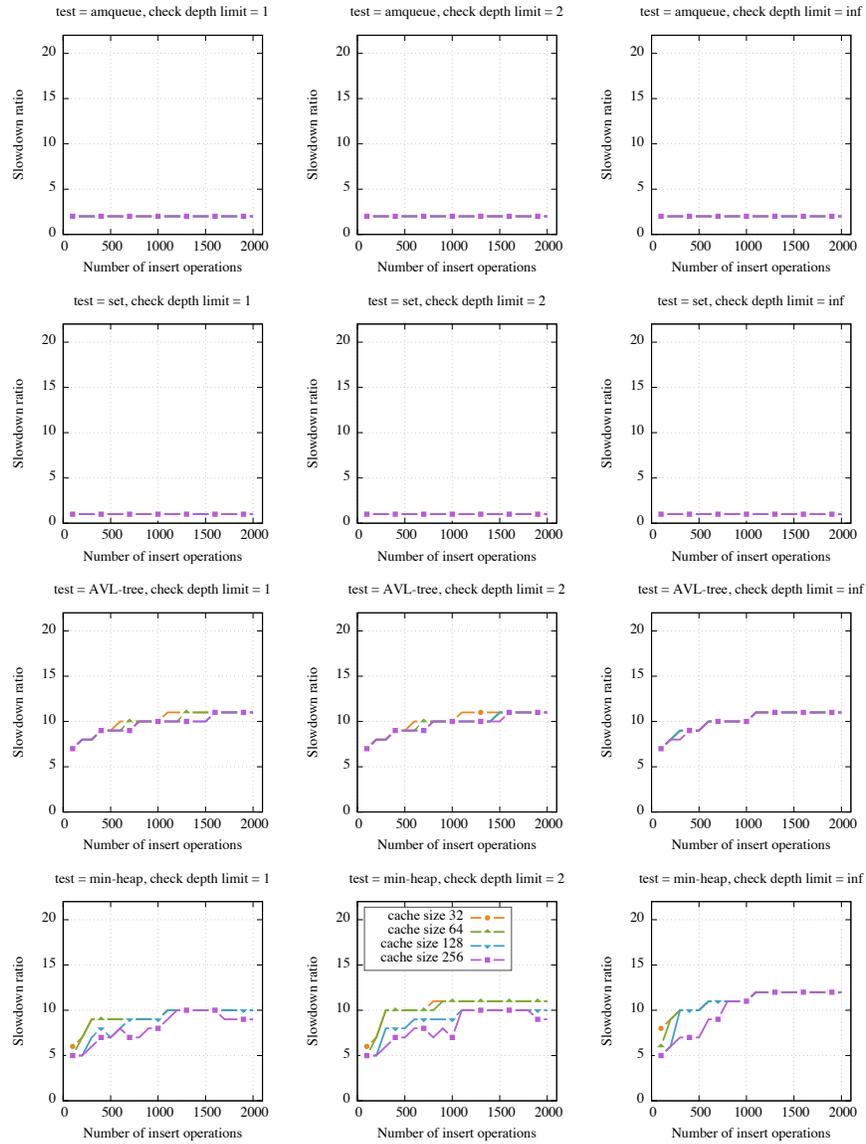

Fig. A 7: Max regtype check depth for each benchmark, check depth limits 1, 2 and $\infty$, DM caching policy.

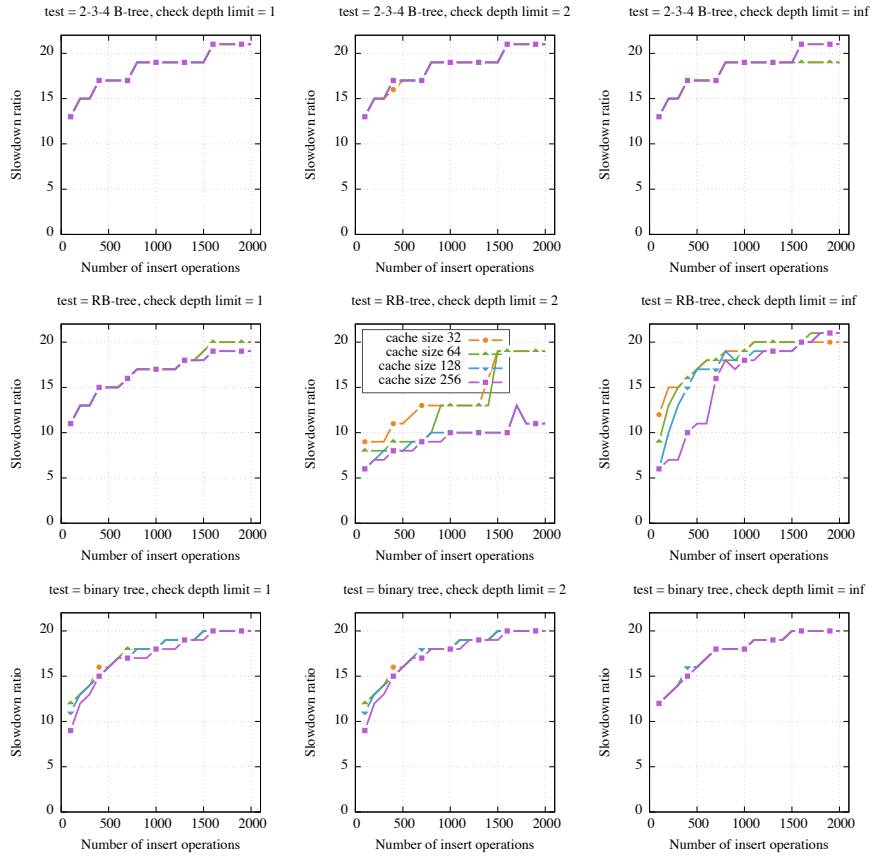

Fig. A 7: Max regtype check depth for each benchmark, check depth limits 1, 2 and ∞, DM caching policy.

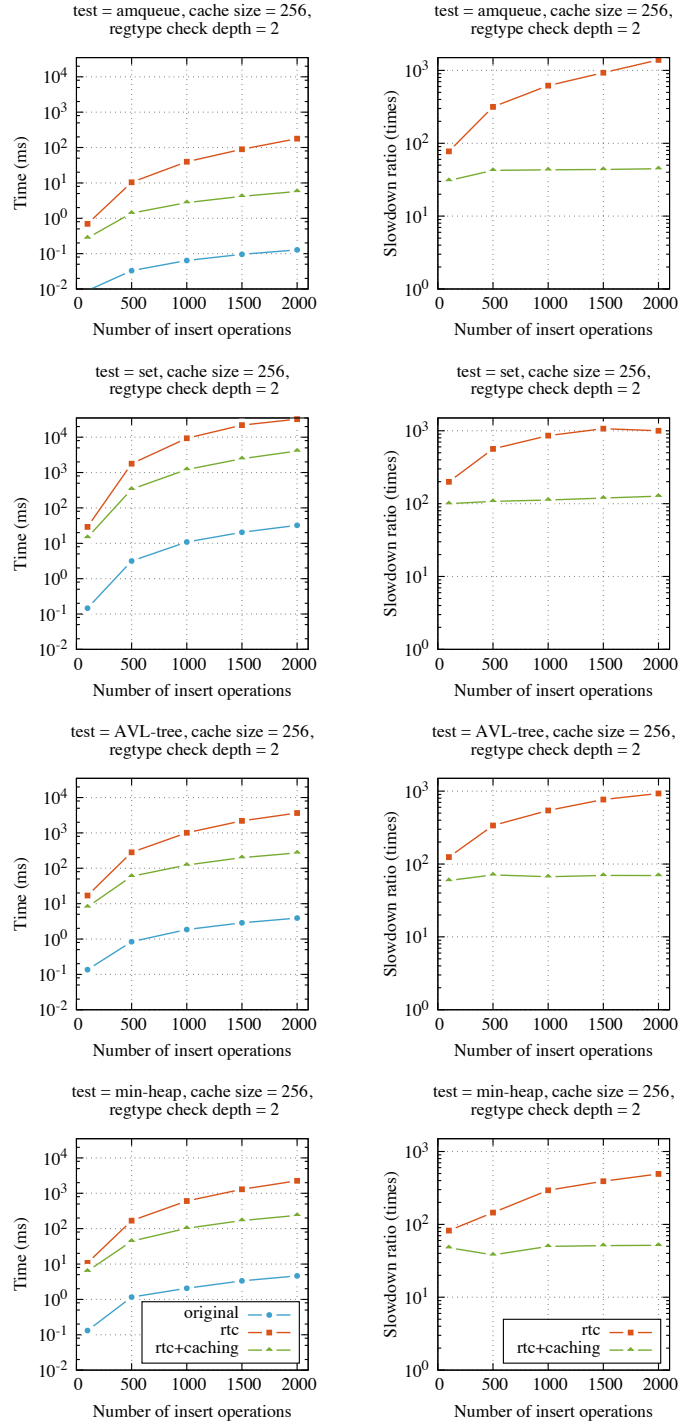

Fig. A 8: Absolute and relative benchmark running times, cache size 256 elements, check depth limit 2, DM caching policy.

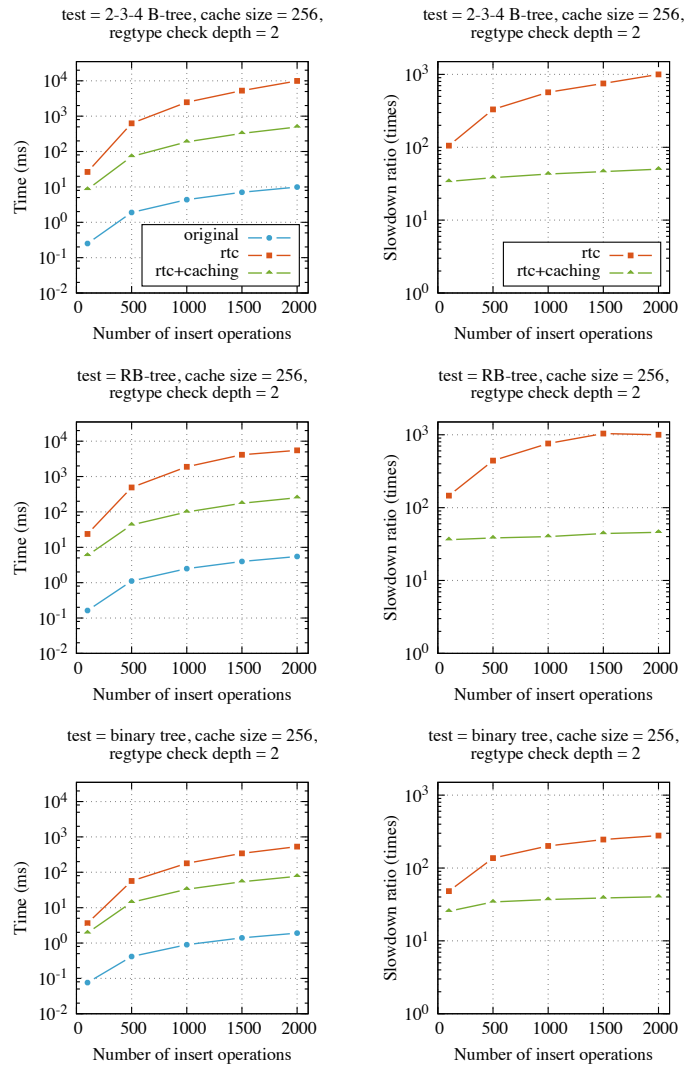

Fig. A 8: Absolute and relative benchmark running times, cache size 256 elements, check depth limit 2, DM caching policy.



\end{document}